\documentclass[]{pasj01}
\draft

\begin{document} 
\Received{}
\Accepted{}

\title{CO Multi-line Imaging of Nearby Galaxies (COMING). VII. Fourier decomposition of molecular gas velocity fields and bar pattern speed}

\author{Dragan \textsc{Salak}\altaffilmark{1}}
\altaffiltext{1}{Department of Physics, School of Science and Technology, Kwansei Gakuin University, 2-1 Gakuen, Sanda, Hyogo 669-1337, Japan}

\author{Yuto \textsc{Noma}\altaffilmark{1}}

\author{Kazuo \textsc{Sorai}\altaffilmark{2,3,4,5}}
\altaffiltext{2}{Department of Physics, Faculty of Science, Hokkaido University, Kita 10 Nishi 8, Kita-ku, Sapporo, Hokkaido 060-0810, Japan}

\author{Yusuke \textsc{Miyamoto}\altaffilmark{6}}

\author{Nario \textsc{Kuno}\altaffilmark{3,4}}
\altaffiltext{3}{Graduate School of Pure and Applied Sciences, University of Tsukuba, 1-1-1 Tennodai, Tsukuba, Ibaraki 305-8571, Japan}
\altaffiltext{4}{Tomonaga Center for the History of the Universe, University of Tsukuba, 1-1-1 Tennodai, Tsukuba, Ibaraki 305-8571, Japan}

\author{Alex R. \textsc{Pettitt}\altaffilmark{2}}

\author{Hiroyuki \textsc{Kaneko}\altaffilmark{7}}

\author{Takahiro \textsc{Tanaka}\altaffilmark{3}}

\author{Atsushi \textsc{Yasuda}\altaffilmark{3}}

\author{Shoichiro \textsc{Kita}\altaffilmark{3}}

\author{Yoshiyuki \textsc{Yajima}\altaffilmark{5}}
\altaffiltext{5}{Department of Cosmosciences, Graduate School of Science, Hokkaido University, Kita 10 Nishi 8, Kita-ku, Sapporo, Hokkaido 060-0810, Japan}

\altaffiltext{6}{National Astronomical Observatory of Japan, 2-21-1 Osawa, Mitaka, Tokyo 181-8588, Japan}

\altaffiltext{7}{Nobeyama Radio Observatory, Minamimaki, Minamisaku, Nagano 384-1305, Japan}

\author{Shugo \textsc{Shibata}\altaffilmark{5}}

\author{Naomasa \textsc{Nakai}\altaffilmark{1,3,4}}

\author{Masumichi \textsc{Seta}\altaffilmark{1}}

\author{Kazuyuki \textsc{Muraoka}\altaffilmark{8}}
\altaffiltext{8}{Department of Physical Science, Osaka Prefecture University, Gakuen 1-1, Sakai, Osaka 599-8531, Japan}

\author{Mayu \textsc{Kuroda}\altaffilmark{8}}

\author{Hiroyuki \textsc{Nakanishi}\altaffilmark{9}}
\altaffiltext{9}{Graduate School of Science and Engineering, Kagoshima University, 1-21-35 Korimoto, Kagoshima, Kagoshima 890-0065, Japan}

\author{Tsutomu T. \textsc{Takeuchi}\altaffilmark{10}}
\altaffiltext{10}{Division of Particle and Astrophysical Science, Nagoya University, Furo-cho, Chikusa-ku, Nagoya, Aichi 464-8602, Japan}

\author{Moe \textsc{Yoda}\altaffilmark{10}}

\author{Kana \textsc{Morokuma-Matsui}\altaffilmark{11}}
\altaffiltext{11}{Institute of Space and Astronautical Science, Japan Aerospace Exploration Agency, 3-1-1 Yoshinodai, Chuo-ku, Sagamihara, Kanagawa 252-5210, Japan}

\author{Yoshimasa \textsc{Watanabe}\altaffilmark{3}}

\author{Naoko \textsc{Matsumoto}\altaffilmark{12}}
\altaffiltext{12}{The Research Institute for Time Studies, Yamaguchi University, Yoshida 1677-1, Yamaguchi, Yamaguchi 753-8511, Japan}

\author{Nagisa \textsc{Oi}\altaffilmark{13}}
\altaffiltext{13}{Faculty of Science Division II, Liberal Arts, Tokyo University of Science, 1-3 Kagurazaka, Shinjuku-ku, Tokyo 162-8601, Japan}

\author{Hsi-An \textsc{Pan}\altaffilmark{14}}
\altaffiltext{14}{Academia Sinica, Institute of Astronomy and Astrophysics, No. 1, Sec. 4, Roosevelt Rd, Taipei 10617, Taiwan}

\author{Ayumi \textsc{Kajikawa}\altaffilmark{15}}
\altaffiltext{15}{Department of Physics, School of Science, Hokkaido University, Kita 10 Nishi 8, Kita-ku, Sapporo, Hokkaido 060-0810, Japan}

\author{Yu \textsc{Yashima}\altaffilmark{15}}

\author{Ryusei \textsc{Komatsuzaki}\altaffilmark{3}}

\email{d.salak@kwansei.ac.jp}


\maketitle

\begin{abstract}
The \(^{12}\)CO (\(J=1\rightarrow0\)) velocity fields of a sample of 20 nearby spiral galaxies, selected from the CO Multi-line Imaging of Nearby Galaxies (COMING) legacy project of Nobeyama Radio Observatory, have been analyzed by Fourier decomposition to determine their basic kinematic properties, such as circular and noncircular velocities. On average, the investigated barred (SAB and SB) galaxies exhibit a ratio of noncircular to circular velocities of molecular gas larger by a factor of 1.5-2 than non-barred (SA) spiral galaxies at radii within the bar semimajor axis \(a_\mathrm{b}\) at 1 kpc resolution, with a maximum at a radius of \(R/a_\mathrm{b}\sim0.3\). Residual velocity field images, created by subtracting model velocity fields from the data, reveal that this trend is caused by kpc-scale streaming motions of molecular gas in the bar region. Applying a new method based on radial velocity reversal, we estimated the corotation radius \(R_\mathrm{CR}\) and bar pattern speed \(\Omega_\mathrm{b}\) in seven SAB and SB systems. The ratio of the corotation to bar radius is found to be in a range of \(\mathcal{R}\equiv R_\mathrm{CR}/a_\mathrm{b}\sim0.8\mathrm{-}1.6\), suggesting that intermediate (SBb-SBc), luminous barred spiral galaxies host fast and slow rotator bars. Tentative negative correlations are found for \(\Omega_\mathrm{b}\) vs. \(a_\mathrm{b}\) and \(\Omega_\mathrm{b}\) vs. total stellar mass \(M_\ast\), indicating that bars in massive disks are larger and rotate slower, possibly a consequence of angular momentum transfer. The kinematic properties of SAB and SB galaxies, derived from Fourier decomposition, are compared with recent numerical simulations that incorporate various rotation curve models and galaxy interactions.
\end{abstract}

\section{Introduction}

The distribution of cold interstellar gas (mainly molecular gas) is closely related to galaxy evolution via its role in star formation (e.g., \cite{Sch11}). In a typical disk-shaped galaxy, molecular gas orbits the dynamical center within a thin disk where most of star formation activity takes place. If the distribution of mass in the disk is non-axisymmetric, in systems such as barred and non-barred spiral galaxies, there are gravitational torques that induce noncircular motions, observable through tracers such as CO and HI.

Molecular gas is an excellent probe of galactic dynamics owing to its low velocity dispersion and highly dissipative nature \citep{SSS80,Sch81,Sch84,Ath92}. For instance, many barred galaxies exhibit large noncircular motions of molecular gas indicative of gas inflows (e.g., \cite{SBS00,Sor00,Sal16}). The phenomenon is understood to be a consequence of shocks in the bar (which propagates as a density wave through the disk), typically offset from the potential minimum, where the frequency of cloud-cloud interactions may be higher than in the rest of the disk (e.g., \cite{Fuj14}). In the bar region, gas clouds lose angular momentum and gradually (on Gyr timescale) inflow toward the galactic center \citep{Ath92}. Consequently, the accumulation of gas in the central region can fuel bursts of star formation and the growth of a pseudobulge linked to secular evolution (e.g., \cite{CE93,KK04,Sel14,Sal17}). To understand the secular evolution of galaxies, it is essential to examine the large-scale kinematics of molecular gas in spiral galaxies and systematically quantify the differences in gas motions between barred and non-barred spiral systems.

On the other hand, in order to investigate the evolution of bars, as prominent non-axisymmetric structures, and their interaction with other components of the host galaxies, such as the dark matter halo, stellar disk, and cold gas, it is important to measure their basic characteristics. Two perhaps most interesting properties of bars from the aspect of galaxy evolution are (1) the bar length \(a_\mathrm{b}\) and (2) the pattern speed \(\Omega_\mathrm{b}\), defined as the angular speed of the bar figure as an elongated ``rigid body'' rotating within the galactic disk.\footnote{The non-axisymmetric potential of a bar has the ability to trap stellar orbits of different radii into a common precession rate (pattern speed), that otherwise tend to precess at different rates if the potential were axisymmetric \citep{LK72,LB79}.} From a theoretical point of view, bars are expected to grow and slow down their rotation over time \citep{Kor13,Sel14}.

The bar length has been studied observationally, e.g., by isophotal analysis (elliptical fitting and mode analysis) of surface brightness structures of near-infrared images that trace the distribution of old stars - the main constituents of bars (e.g., \cite{EE85,HE15,Sal15,DG16}). On the other hand, the bar pattern speed, as well as the pattern speed of spiral arms (both often regarded as density waves), have been estimated usually from the velocity fields of stars and interstellar gas. Some examples of the often applied methods include the Tremaine-Weinberg method \citep{TW84a,MK95,Agu15,Guo19}, which utilizes a tracer that obeys the continuity equation (e.g., old stars), potential-density phase-shift method \citep{ZB07}, various measurements using molecular gas data (e.g., \cite{Kun00,Kod02,Hir14}), the Font-Beckman method \citep{Fon11,Fon17}, and others. While recent studies have yielded measurements of \(\Omega_\mathrm{b}\) in an increasing number of nearby galaxies, the importance of measuring this quantity and investigating its relation with fundamental galactic parameters, such as the total stellar mass, have motivated us to make new estimates of \(\Omega_\mathrm{b}\) using \(^{12}\)CO (\(J=1\rightarrow0\)) data as a well-established tracer of molecular gas. In this paper, we report two major results of an analysis of molecular gas velocity fields of 20 nearby spiral galaxies, including a subsample of 7 barred galaxies: (1) new measurements of basic galactic parameters, such as the position angle, inclination, and systemic velocity, and (2) kinematic parameters in terms of circular and noncircular velocity components and bar pattern speed.

The structure of the paper is as follows. In section \ref{B}, we describe the analysis and the sample data. The results, including the derived basic galactic parameters, circular and noncircular velocities, are presented in section \ref{C}, followed by a discussion on the bar pattern speed from the aspect of observations and numerical simulations (section \ref{D}) and summary (section \ref{E}).

\section{Data analysis}\label{B}

We begin by describing the basic principles behind the mode analysis (Fourier decomposition) of the CO (1-0) velocity fields, that was applied in this work, and the tools that were used to perform the computation.

\subsection{Velocity field as Fourier series}\label{B1}

\begin{figure}
\begin{center}
\includegraphics[width=16cm]{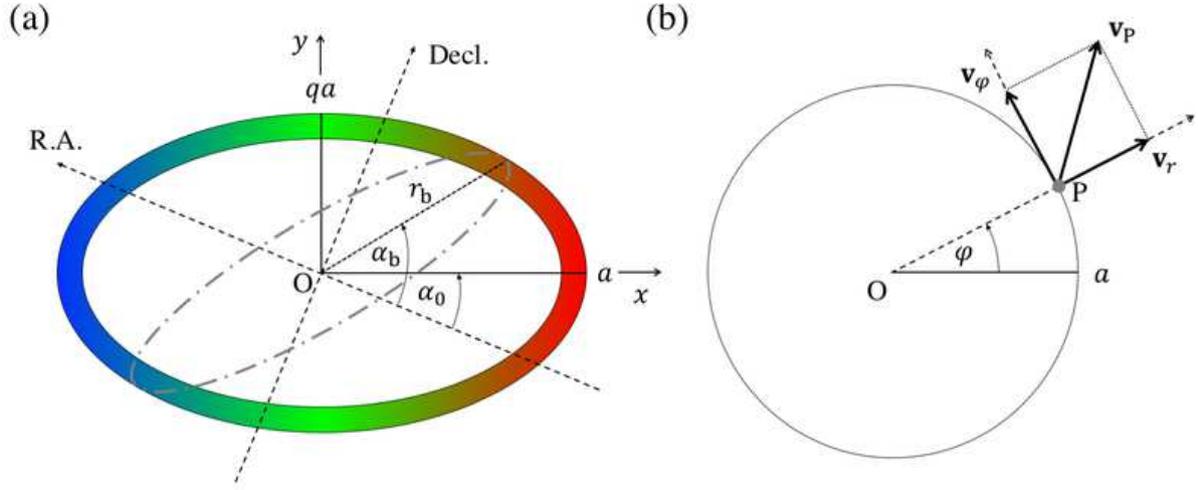}
\end{center}
\caption{(a) Illustration of the geometry of an elliptical ring at a position angle \(\alpha_0\) with respect to the sky coordinates (R.A., Decl). The semimajor and semiminor axes are \(a\) and \(qa\), respectively. The color gradient indicates the change in velocity from redshift to blueshift. The standard position angle, defined as the angle of the redshifted side measured anti-clockwise from north, in this case is \(\mathrm{PA}=270\arcdeg+\alpha_0\). The angle \(\varphi\) (eccentric anomaly) in equation \ref{FS} is measured anti-clockwise from \(x\)-axis. In the case of a barred galaxy, \(r_\mathrm{b}\) is the projected bar radius and \(\alpha_\mathrm{b}\) is its position angle (section \ref{C3}). The bar is illustrated with a dot-dashed grey ellipse. (b) Definitions of radial velocity \(v_r\) and azimuthal velocity \(v_\varphi\) at point P in the reference frame of a disk.}
\label{fig:ell}
\end{figure}

The velocity field of a rotating disk galaxy (barred or non-barred) can be represented as a set of elliptical rings centered at the position of the dynamical center.\footnote{A ring is assumed to be circular but inclined with respect to the observer so that it appears as an ellipse.} If \(a\) is the semimajor axis of an ellipse and \(\varphi\) is the azimuthal angle in polar coordinates measured from the semimajor axis, then the velocity field \(v= v(a,\varphi)\) sampled on any such elliptical ring is a periodic function of \(\varphi\) and can be expressed as a Fourier series

\begin{equation}\label{FS}
v(a,\varphi)=c_0+\sum_{m=1}^\infty c_m\cos(m\varphi)+\sum_{m=1}^\infty s_m\sin(m\varphi),
\end{equation}
and the coordinates of the ellipse are given by \(x_\mathrm{e}=a\cos{\varphi}\) and \(y_\mathrm{e}=qa\sin{\varphi}\), where \(q\) is the ratio of the minor to major axis, also called flattening (see figure \ref{fig:ell}). Note that, in the case that \(m=1\) is the only mode, equation \ref{FS} becomes \(v=c_0+c_1\cos{\varphi}+s_1\sin{\varphi}\). If we define \(c_0\equiv v_\mathrm{sys}\), \(c_1\equiv v_\varphi\sin{i}\), and \(s_1\equiv v_r\sin{i}\), the above equation can be written as \(v=v_\mathrm{sys}+v_\varphi\sin{i}\cos{\varphi}+v_r\sin{i}\sin{\varphi}\), where \(v_\mathrm{sys}\) is the systemic velocity, \(i\) is the inclination angle of the plane of the galactic disk, \(v_r\) is the radial component, and \(v_\varphi\) is the circular component of the velocity vector in the disk, as defined in figure \ref{fig:ell}(b). This equation describes the planar motion of gas in a thin disk, and the coefficients \(s_1\) and \(c_1\), respectively, correspond to the axisymmetric radial and circular velocity components in the disk observed at an inclination angle \(i\). The sign of \(s_1\) can be used as an indicator of axisymmetric inward or outward motions if the orientation of the disk (near/far side) is known. In a galaxy with purely circular rotation, the velocity field is given by \(v=v_\mathrm{sys}+c_1\cos{\varphi}\) and \(c_1/\sin{i}\) is equivalent to the rotation curve. If noncircular motions are present, their axisymmetric component is described by \(s_1\), while nonaxisymmetric components are included in higher order coefficients \(m\geq2\) \citep{SFZ97}. The higher coefficients, in general, provide indications of local irregular motions in the disk, gravitational perturbations of order higher than 2 \citep{Mei08}, and may arise from errors in the galactic center position and geometric parameters \citep{Kra06}. For rotating disk galaxies, it is reasonable to expect that \(c_1\) and \(s_1\) are the dominant terms at most radii (section \ref{C}). Note that all coefficients are functions of semimajor axis only.

In the present analysis, by ``Fourier decomposition'' we refer to a method of calculating the coefficients of equation \ref{FS} up to \(m=3\) from a given velocity field data set.

\subsection{The data}\label{B2}

The data used in the analysis consist of \(^{12}\)CO (1-0) images acquired from the CO Multi-line Imaging of Nearby Galaxies (COMING), which is a legacy project of Nobeyama Radio Observatory \citep{Sor19}. On-the-fly mapping observations of 147 nearby galaxies in the \(^{12}\)CO, \(^{13}\)CO, and C\(^{18}\)O (1-0) lines were carried out using the FOur beam REceiver System on the 45-m Telescope [FOREST; \citet{Min16}]. The acquired \(^{12}\)CO (1-0) data have an effective spatial resolution of \(17\arcsec\), image grid spacing of \(d=6\arcsec\), and a velocity resolution of \(\Delta v_\mathrm{ch}=10~\mathrm{km~s^{-1}}\). Although the distances to the galaxies in the sample range from \(4\) to \(39\) kpc (table \ref{tab:gal}), the average distance yields a physical resolution of \(\sim1.6~\mathrm{kpc}\); thus, the results of the present study hold on a kiloparsec scale. A detailed description of the observations and data reduction, as well as a gallery of CO (1-0) images and star formation properties of all sample galaxies, are given in the forthcoming papers by \citet{Sor19} (COMING overview), \citet{Mur19}, and \citet{Yaj19}. Throughout this paper, CO refers to the \(^{12}\)CO isotopologue.

In the data analysis, we used moment 1 images (two-dimensional data), defined as intensity-weighted velocity \(\langle v\rangle\equiv\int V T_\mathrm{mb}dV/\int T_\mathrm{mb}dV\), where \(T_\mathrm{mb}\) is the main beam brightness temperature and \(V\) is the velocity with respect to the local standard of rest (LSR) in radio definition. The data cubes were clipped at \(4\sigma\) in \(T_\mathrm{mb}\) prior to creating moment 1 images, where the sensitivity was typically better than \(1\sigma(T_\mathrm{mb})\approx80~\mathrm{mK}\).

The criteria for selecting galaxies for the present analysis were the following: (1) \(^{12}\)CO (1-0) emission detected at \(\geq4\sigma\) in \(T_\mathrm{mb}\) within a field of radius of at least \(R\sim50\arcsec\) (\(\sim3\times\) effective beam size), (2) within a radius larger than the projected bar radius (in barred galaxies), and (3) not edge-on (inclination \(i\leq75\arcdeg\)). These conditions allowed us to perform Fourier decomposition of velocity field images, as described in the next section, and derive basic galactic properties, such as position angle, inclination, systemic velocity, and circular/noncircular velocities. The sample of 20 galaxies, listed in table \ref{tab:gal}, includes all major types of spiral galaxies classified by the presence or absence of a bar according to \citet{deV91}: SA (non-barred), SAB (weakly barred), and SB (strongly barred) systems.

\subsection{Fourier decomposition of CO (1-0) velocity fields}\label{B3}

To calculate the coefficients in equation \ref{FS} from CO (1-0) moment 1 images, we used Kinemetry \citep{Kra06}, a program that fits elliptical rings on an observed velocity field \(\langle v\rangle\) by minimizing the noncircular velocity component expressed as \(v_\mathrm{nc}=\sqrt{s_1^2+s_2^2+c_2^2+s_3^2+c_3^2}\) (up to \(m=3\)). This is done by searching for a minimum value of \(\langle v\rangle-c_1\cos{\varphi}\) (residuals from circular rotation models) on a grid of flattenings \(0.2\leq q\leq1\) and kinematic position angles \(-90\degree\leq\alpha_0\leq90\degree\). The flattening is related to the galactic inclination (assuming a thin disk) as \(q=\cos{i}\). For each ring, a pair (\(q,\alpha_0\)) with the minimum \(v_\mathrm{nc}\) yields the parameters of the best-fit ellipse, whose coordinates are \((x_\mathrm{e},y_\mathrm{e})\). The data points in incomplete rings are retrieved by interpolation and the calculation is performed until a ring which is filled with less than 75\% of the data points.

\begin{figure}
\begin{center}
\includegraphics[width=16cm]{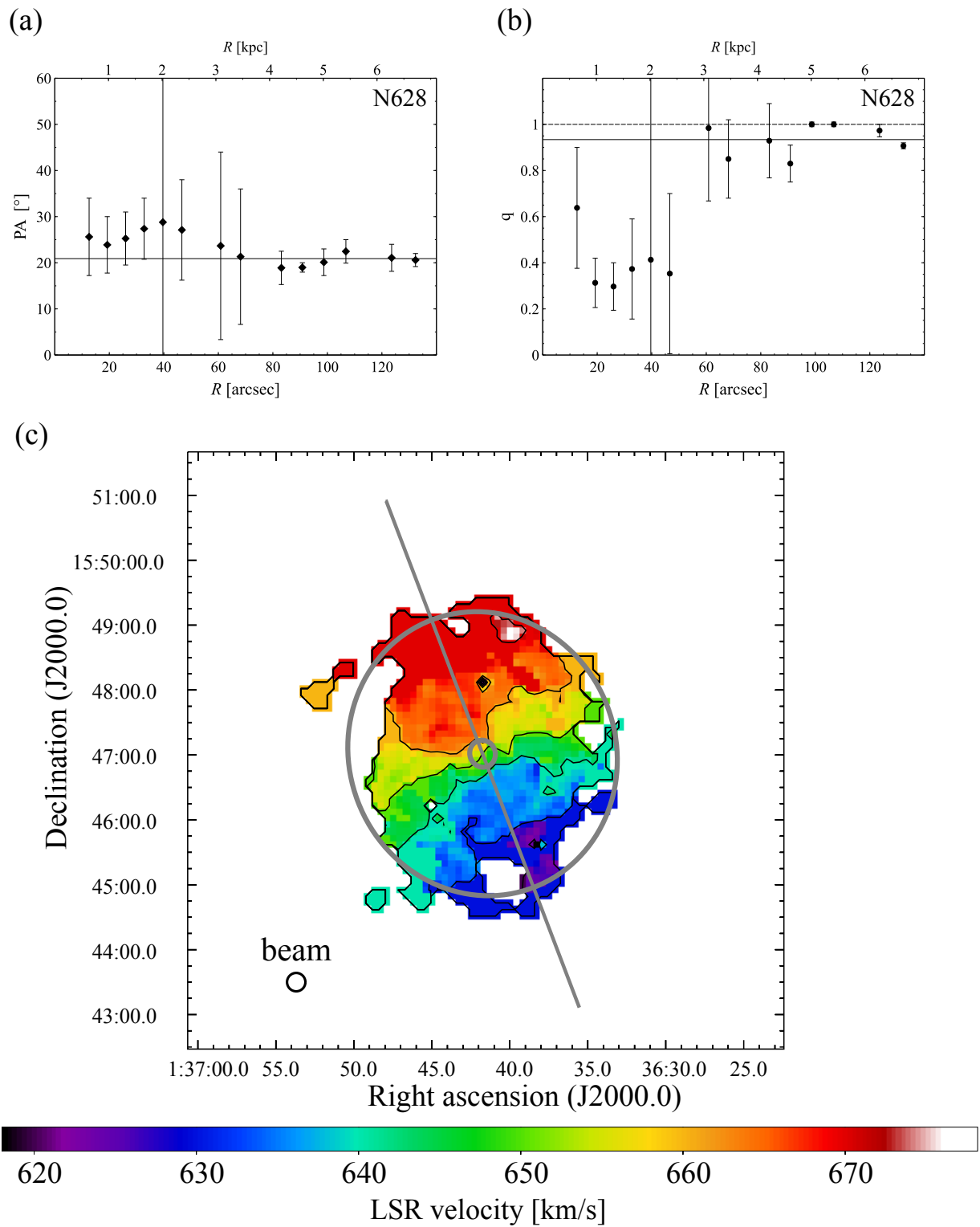}
\end{center}
\caption{(a) Position angle (PA), and (b) flattening \(q\) calculated for NGC 628. The horizontal full lines show the weighted average values calculated for rings beyond \(R>50\arcsec\) (table \ref{tab:gal}). The horizontal dashed line in panel (b) marks \(q=1\). (c) Moment 1 image of NGC 628. The contours are plotted at \(\langle v\rangle=631,641,651,661,671~\mathrm{km~s^{-1}}\); the straight line indicates the derived kinematic major axis, i.e., position angle \(\langle PA\rangle\) (table \ref{tab:gal}). Also shown are the innermost (\(a=12.6\arcsec\)) and the outermost (\(a=132.3\arcsec\)) elliptical rings calculated using \(\langle PA\rangle\) and \(\langle q\rangle\).}
\label{fig:n628}
\end{figure}

The program was first set to calculate the initial values of \(\alpha_0\) and \(q\) for each ring. The galactic center coordinates (table \ref{tab:gal}) were fixed, and the semimajor axes of the rings were determined by the program's default sampling function \(a_n=(n+1.1^n)d\), where \(n=0,1,2,...,N\) is the ring number, and \(d=6\arcsec\) is the grid spacing on images. The initial values of \(\alpha_0\) and \(q\) were derived for each radius (figure \ref{fig:n628}). The mean values of the two parameters across the disk of each galaxy were calculated as \(\langle\alpha_0\rangle=\sum_n^N w_n \alpha_{0n}/\sum_n^N w_n\) and \(\langle q\rangle=\sum_n^N w_n q_n/\sum_n^N w_n\), where \(w=w(a)\) is a weighting function equal to the number of ellipses used to determine the best fit within each ring \(n\), and \(N\) is the total number of rings for which fitting was successful. Applying this function ensures that rings with a larger number of data points, which are found in the outer parts of images, are given more weight in calculating the average values. Here, we assume that the CO gas disks are not significantly warped within the investigated regions. The innermost ring and the outermost ring were excluded from calculations because the number of pixels and the signal-to-noise ratio were minimal, respectively. In addition, we found that the initial values of \(q\) were notably different in the inner \(R<50\arcsec\) region compared to the rest of the disk in NGC 628 (figure \ref{fig:n628}), and within \(R<40\arcsec\) in NGC 3893. For these two galaxies, only rings beyond these radii were used in calculating \(\langle q\rangle\) and \(\langle\alpha_0\rangle\). For barred galaxies, the weighted averages were calculated using only the values of the rings outside the bar region, defined by the bar length in table \ref{tab:gal1}. This is justified by the expectation that noncircular motions are dominant in the bar region, and may induce an ``S''-shaped disturbance of the velocity field that resembles a warp of the disk. Therefore, taking the average over all rings in the disk would yield an average kinematic position angle of that of the inner and outer regions of the disk. We assume that the outer region is less affected by the tangential forces induced by the bar, so that the kinematics of molecular gas there is well represented by circular rotation. As shown in section \ref{C1}, this assumption is reasonable given the excellent agreement between the position angles derived by Kinemetry and those by photometric methods for both barred and non-barred spirals. The calculated weighted average values (table \ref{tab:gal}) were fixed for all radii in the next round of calculations that yielded the final results. The output parameters are the Fourier coefficients \(c_m\) and \(s_m\) (\(m=1,2,3\)), kinematic position angle \(\langle\alpha_0\rangle\), flattening \(\langle q\rangle\), and systemic velocity \(v_\mathrm{sys}(\equiv c_0)\). The systemic velocity was calculated as the mean of the values determined in rings that were used to calculate \(\langle\alpha_0\rangle\) and \(\langle q\rangle\).

\begin{table}
  \tbl{Basic parameters of the sample galaxies}{
  \begin{tabular}{lccccccccc}
      \hline
      Galaxy & \(\alpha_\mathrm{J2000.0}\) & \(\delta_\mathrm{J2000.0}\) & Morphology & Distance & \(\langle \alpha_0\rangle\) & \(\langle\mathrm{PA}\rangle\) & \(\langle q\rangle\) & \(\langle i\rangle\) & \(v_\mathrm{sys}\) \\
       & \(\mathrm{(h:m:s)}\) & \(\mathrm{(d:m:s)}\) & & (Mpc) & \((\degree)\) & \((\degree)\) & \((\degree)\) & \((\degree)\) & \((\mathrm{km~s}^{-1})\) \\ 
      NGC (Messier) & (1) & (1) & (2) & (3) & (4) & (5) & (6) & (7) & (8) \\
      \hline
      157 & \(00:34:46.76\) & \(-08:23:47.2\) & SAB(rs)bc & \(12.1\) & \(-43.5\pm1.2\) & \(226.5\pm1.2\) & \(0.536\pm0.083\) & \(57.6\pm5.6\) & \(1641.0\pm1.8\) \\
      613 & \(01:34:18.170\) & \(-29:25:06.10\) & SB(rs)bc & \(26.4\) & \(27.0\pm0.4\) & \(297.0\pm0.4\) & \(0.405\pm0.025\) & \(66.1\pm1.6\) & \(1477.7\pm1.3\) \\
      628 (M74) & \(01:36:41.747\) & \(+15:47:01.18\) & SA(s)c & \(9.02\) & \(-69.1\pm3.4\) & \(20.9\pm3.4\) & \(0.934\pm0.057\) & \(20.9\pm9.1\) & \(651.1\pm1.3\) \\
      1084 & \(02:45:59.908\) & \(-07:34:42.48\) & SA(s)c & \(20.9\) & \(-58.0\pm0.5\) & \(212.0\pm0.5\) & \(0.703\pm0.089\) & \(45.3\pm7.2\) & \(1402.2\pm3.6\) \\
      2903 & \(09:32:10.11\) & \(+21:30:03.0\) & SAB(rs)bc & \(9.46\) & \(-64.9\pm0.8\) & \(205.1\pm0.8\) & \(0.258\pm0.038\) & \(75.0\pm2.3\) & \(540.7\pm0.9\)  \\
      2967 & \(09:42:03.295\) & \(+00:20:11.18\) & SA(s)c & 22.2 & \(35.3\pm1.5\) & \(305.3\pm1.5\) & \(0.62\pm0.15\) & \(52\pm11\) & \(1875.0\pm5.2\) \\
      3147 & \(10:16:53.65\) & \(+73:24:02.695\) & SA(rs)bc & \(39.3\) & \(56.1\pm0.7\) & \(146.1\pm0.7\) & \(0.908\pm0.028\) & \(24.8\pm3.8\) & \(2773.0\pm2.7\) \\
      3627 (M66) & \(11:20:14.964\) & \(+12:59:29.54\) & SAB(s)b & \(9.04\) & \(76.4\pm1.5\) & \(166.4\pm1.5\) & \(0.535\pm0.044\) & \(57.7\pm3.0\) & \(704.6\pm1.8\) \\
      3810 & \(11:40:58.760\) & \(+11:28:16.10\) & SA(rs)c & \(16.4\) & \(-58.9\pm1.2\) & \(211.1\pm1.2\) & \(0.891\pm0.065\) & \(27.0\pm8.2\) & \(992.9\pm2.7\) \\
      3893 & \(11:48:38.19\) & \(+48:42:39.0\) & SAB(rs)c & \(15.7\) & \(75.6\pm3.6\) & \(345.6\pm3.6\) & \(0.789\pm0.032\) & \(37.9\pm3.0\) & \(980.7\pm1.3\) \\
      3938 & \(11:52:49.45\) & \(+44:07:14.6\) & SA(s)c & \(17.9\) & \(-67.7\pm1.1\) & \(202.3\pm1.1\) & \(0.951\pm0.042\) & \(18.0\pm7.8\) & \(812.2\pm0.6\) \\
      4030 & \(12:00:23.627\) & \(-01:06:00.34\) & SA(s)bc & \(29.9\) & \(-53.8\pm0.5\) & \(36.2\pm0.5\) & \(0.933\pm0.045\) & \(21.1\pm7.2\) & \(1458.7\pm2.5\) \\
      4303 (M61) & \(12:21:54.895\) & \(+04:28:25.13\) & SAB(rs)bc & \(16.5\) & \(45.3\pm0.4\) & \(315.3\pm0.4\) & \(0.942\pm0.031\) & \(19.6\pm5.3\) & \(1562.6\pm1.3\) \\
      4579 (M58) & \(12:37:43.522\) & \(+11:49:05.498\) & SAB(rs)b & \(16.5\) & \(0.3\pm0.7\) & \(90.3\pm0.7\) & \(0.780\pm0.040\) & \(38.7\pm3.7\) & \(1507.6\pm1.8\) \\
      5055 (M63) & \(13:15:49.33\) & \(+42:01:45.4\) & SA(rs)bc & \(9.04\) & \(8.8\pm0.3\) & \(98.8\pm0.3\) & \(0.506\pm0.017\) & \(59.6\pm1.1\) & \(498.5\pm2.3\) \\
      5248 & \(13:37:32.024\) & \(+08:53:06.64\) & SAB(rs)bc & \(13.0\) & \(12.7\pm1.3\) & \(102.7\pm1.3\) & \(0.874\pm0.044\) & \(29.1\pm5.2\) & \(1151.7\pm1.4\) \\
      5676 & \(14:32:46.846\) & \(+49:27:28.45\) & SA(rs)bc & \(34.7\) & \(-47.1\pm0.6\) & \(222.9\pm0.6\) & \(0.735\pm0.015\) & \(42.7\pm1.3\) & \(2116.8\pm8.2\) \\
      5678 & \(14:32:05.610\) & \(+57:55:17.20\) & SAB(rs)b & \(35.7\) & \(-83.2\pm0.7\) & \(186.8\pm0.7\) & \(0.731\pm0.093\) & \(43.0\pm7.8\) & \(1899.0\pm7.5\) \\
      6643 & \(18:19:46.41\) & \(+74:34:06.1\) & SA(rs)c & \(21.3\) & \(-50.0\pm0.7\) & \(30.0\pm0.7\) & \(0.629\pm0.033\) & \(51.0\pm2.4\) & \(1496.2\pm2.1\) \\
      7479 & \(23:04:56.65\) & \(+12:19:22.4\) & SB(s)c & \(36.8\) & \(-62.7\pm1.8\) & \(207.3\pm1.8\) & \(0.495\pm0.061\) & \(60.3\pm4.0\) & \(2346.8\pm5.5\) \\
      \hline
\end{tabular}}\label{tab:gal}
\begin{tabnote}
(1) Adopted from NASA/IPAC Extragalactic Database. (2) \citet{deV91}. (3) NGC 613: \citet{Tul09}, NGC 628: \citet{Dhu16}, NGC 2967: \citet{Bot84}, NGC 3893: \citet{Spr09}, NGC 3938: \citet{Poz09}, NGC 4303 and NGC 4579: \citet{Mei07}, others: \citet{Tul13}. (4) The derived kinematic position angle \(\langle\alpha_0\rangle\) (measured anti-clockwise from west) is also expressed as (5) the position angle of the receding side of the galaxy \(\langle\mathrm{PA}\rangle\), which is measured anti-clockwise from north. (6) Ellipse flattening \(q=\cos{i}\). (7) Galactic disk inclination. (8) The velocities are in radio definition with respect to the local standard of rest (LSR).
\end{tabnote}
\end{table}

\section{Results}\label{C}

In this section, we describe the main results of the analysis using Kinemetry, presented for each galaxy in terms of geometric and kinematic parameters, circular and radial velocity components, and the fraction of noncircular motions as a function of galactocentric radius.

\subsection{Basic geometric and kinematic parameters}\label{C1}

\begin{figure}
\begin{center}
\includegraphics[width=17cm]{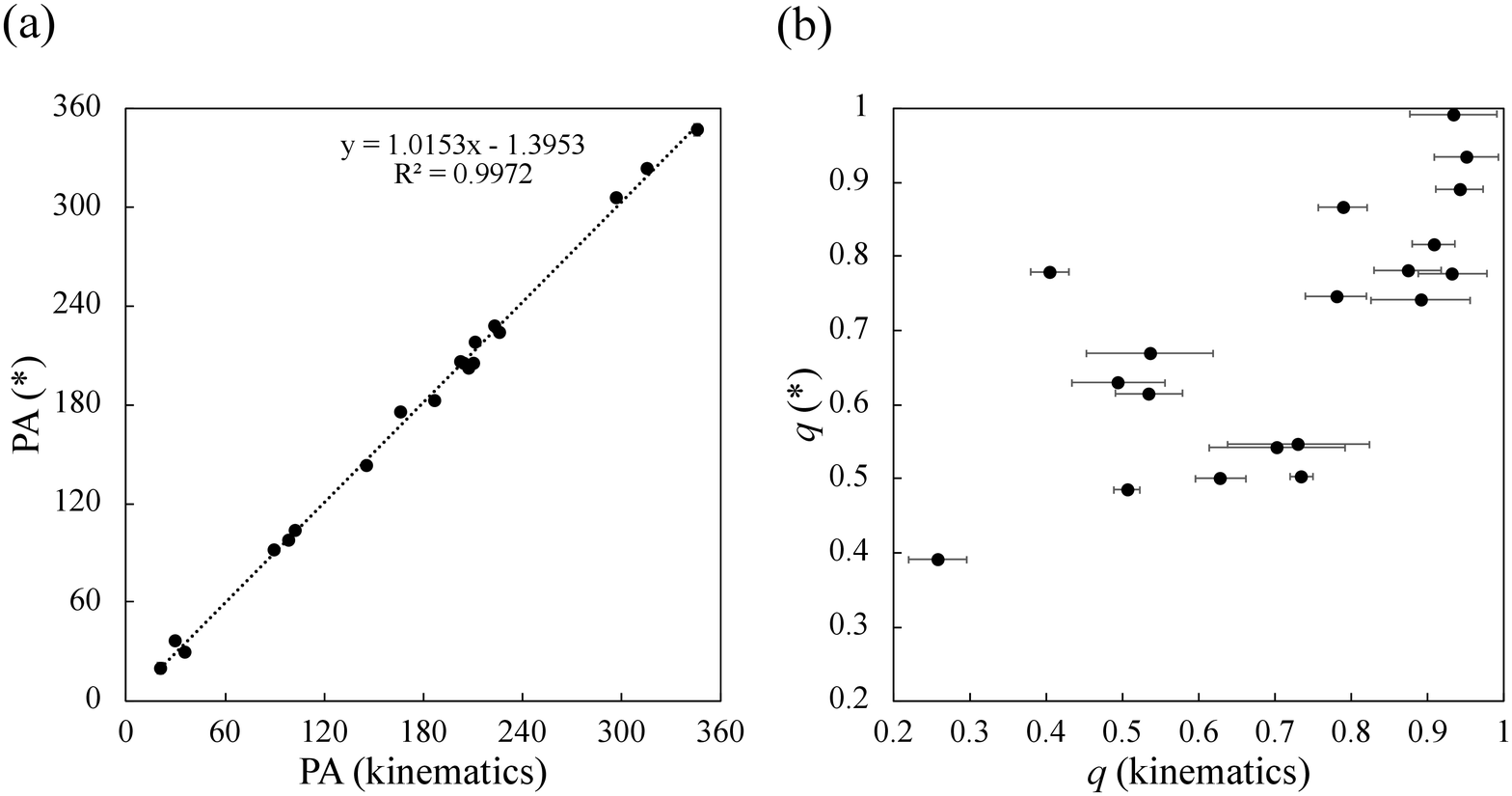} 
\end{center}
\caption{Comparison of (a) position angle PA and (b) flattening \(q\) derived from kinematics in this work with the literature values compiled in \citet{Sor19}, denoted by (*). The uncertainties are approximately the size of data markers in panel (a).}
\label{fig:pa-incl}
\end{figure}

As described in the previous section, the program fitted elliptical rings on sampled moment 1 images. The solutions yielded several basic kinematic parameters, such as (1) the weighted-average of the kinematic position angle \(\langle\alpha_0\rangle\) of the galactic disk (ellipse), measured within \((-90\arcdeg,+90\arcdeg)\) anti-clockwise from west, (2) the flattening \(\langle q\rangle\), and (3) the systemic velocity \(v_\mathrm{sys}\), which is the zeroth-term Fourier coefficient \(c_0\). The obtained parameters are listed in table \ref{tab:gal}; a comparison with the values of position angle (PA) and \(q\), compiled from the literature in \citet{Sor19} and derived mainly from photometry, is shown in figure \ref{fig:pa-incl}. In the case of NGC 2967, the photometric value of \(\mathrm{PA}=64.0\arcdeg\pm62.5\arcdeg\) \citep{Sal15} was largely offset from our value of \(\langle\mathrm{PA}\rangle=305.3\arcdeg\pm1.5\arcdeg\) determined from kinematics and so was excluded from the plots in figure \ref{fig:pa-incl}. The photometric PA was obtained from the analysis of the Survey of Stellar Structure in Galaxies (\(\mathrm{S^4G}\)) \emph{Spitzer} \(3.6~\micron\) images. A comparison of the photometric PA as a function of radius in \citet{Sal15} with \(\langle\mathrm{PA}\rangle\) in table \ref{tab:gal} suggests that the two values are consistent in the inner \(R\lesssim40\arcsec\) of this galaxy, whereas the photometric PA exhibits a wide scatter at larger radii. The linear relation between photometric and kinematic PAs for the galaxies in figure \ref{fig:pa-incl}(a) is very tight, which is important given that the sample includes both barred and non-barred spirals; the consistency for the barred systems supports our assumption made in section \ref{B3} that the kinematic PA in such systems is more reliably determined if only rings outside the bar region are considered.

Note that the relation between the kinematic and photometric estimates exhibits a much larger scatter in the case of \(q\), as evident from figure \ref{fig:pa-incl}(b). The cause of this is not clear, but could be an effect of disk warps at large radii that affect photometric measurements. Many galactic disks are warped in the outermost regions; \(q\) determined in the outer disk is thus not necessarily the same as the one in the inner disk, where we determined the kinematic \(q\) from CO velocity fields. For example, our estimate of \(\langle q\rangle=0.934\pm0.057\) for NGC 628 gives \(\langle i\rangle=21\arcdeg\pm9\arcdeg\), compared to some previous works that suggest an inclination as small as \(7\arcdeg\) in the inner disk \citep{KB92}. On the other hand, our measurement for this galaxy is in good agreement with \(i\approx20\arcdeg\) derived from optical data in HyperLeda\footnote{http://leda.univ-lyon1.fr/} \citep{Pat03}. Compared to the mean inclinations derived from The HI Nearby Galaxy Survey (THINGS) data in \citet{dB08} and \citet{Tra08}, the differences between HI and our CO measurements for four galaxies that were analyzed in both projects are: \(\sim1\arcdeg\) in NGC 5055, \(\sim3\arcdeg\) in NGC 3627, \(\sim6\arcdeg\) in NGC 628 (within uncertainties), and \(\sim9\arcdeg\) in NGC 2903. The inclination of NGC 2903, derived from HI as a function of radius in these works, is closer to our weighted average (\(\sim75\arcdeg\)) in the inner \(R\lesssim100\arcsec\), but decreases at large radii, yielding an average of \(\sim66\arcdeg\). Given the differences in the angular resolution and properties of HI and CO gases, the results in the two studies are in reasonably good agreement.

\subsection{Circular and radial velocities}\label{C2}

The deprojected azimuthal velocity, defined as \(v_\varphi=c_1/\sin{i}\), where \(c_1\) is the amplitude of the first-order cosine term and \(i\) is the inclination of the galactic disk, was calculated for a number of radii determined by the size of the region in the moment 1 map where CO (1-0) emission was detected at \(\geq4\sigma\). The derived \(v_\varphi\) was used to obtain the angular velocity, \(\Omega(R)=v_\varphi(R)/R\), and these two quantities describe the circular motion of molecular gas in a galactic disk.

The program also calculated the deprojected radial velocity component, \(v_r=s_1/\sin{i}\), where \(s_1\) is the amplitude of the first-order sine term. This velocity corresponds to axisymmetric inward or outward motion corrected for galactic inclination. The term is expected to be non-zero, for example, in a system where noncircular motions of gas arise from the potential of a bar. The ratio of overall noncircular to circular motions in a galactic disk is quantified by

\begin{equation}
\Delta(R)\equiv \frac{v_\mathrm{nc}}{c_1},
\end{equation}
where \(v_\mathrm{nc}=\sqrt{s_1^2+s_2^2+c_2^2+s_3^2+c_3^2}\) is a measure of noncircular motions as a function of radius (section \ref{B3}).

Examples of the results of calculations of \(v_\mathrm{\varphi}\), \(\Omega\), \(v_r\), and \(\Delta\) for two representative galaxies of SA (NGC 5055) and SB (NGC 2903) types are shown in figure \ref{fig:ex}, while the rest of the sample is plotted in Appendix. Figure \ref{fig:ex} shows several important features of these systems:

(1) The kinematics is clearly dominated by circular rotation at all radii (\(v_r\ll v_\varphi\)), especially outside the bar region, and the derived \(v_\varphi\) of both galaxies are consistent with the shapes and amplitudes of the rotation curves derived from HI data \citep{dB08}.

(2) The fraction of noncircular motions is somewhat larger in the inner region of the barred galaxy; within \(R<50\arcsec\), we find \(\Delta\sim0.15\) in NGC 5055 and \(\Delta\sim0.20\) in NGC 2903. A comparison with \(v_r/v_\varphi\) suggests that \(\Delta\) is dominated by the \(s_1\) term, i.e., axisymmetric radial flow. As shown below, this behavior is a general property of molecular gas motion in all investigated barred galaxies. While noncircular motions are not entirely absent in NGC 5055, they are a factor of 1.5-2 smaller than in NGC 2903; noncircular motions in this SA galaxy are possibly caused by gravitational torques due to the spiral arms.

(3) Panel (c) shows a nearly linear rise of \(v_\varphi\) in NGC 2903 up to the radius \(R\sim4~\mathrm{kpc}\), resembling the rotation of a rigid body. This is a consequence of the fact that the bar position angle of this galaxy is nearly the same as the position angle of the host galaxy (the difference is \(\theta\sim3\arcdeg\); see table \ref{tab:gal1}). As discussed in section \ref{D13}, the observed circular velocity in this case traces the rotation of the bar itself and reveals its angular velocity, which appears to be nearly constant within \(30\arcsec<R<90\arcsec\).

\begin{figure}
\begin{center}
\includegraphics[width=17cm]{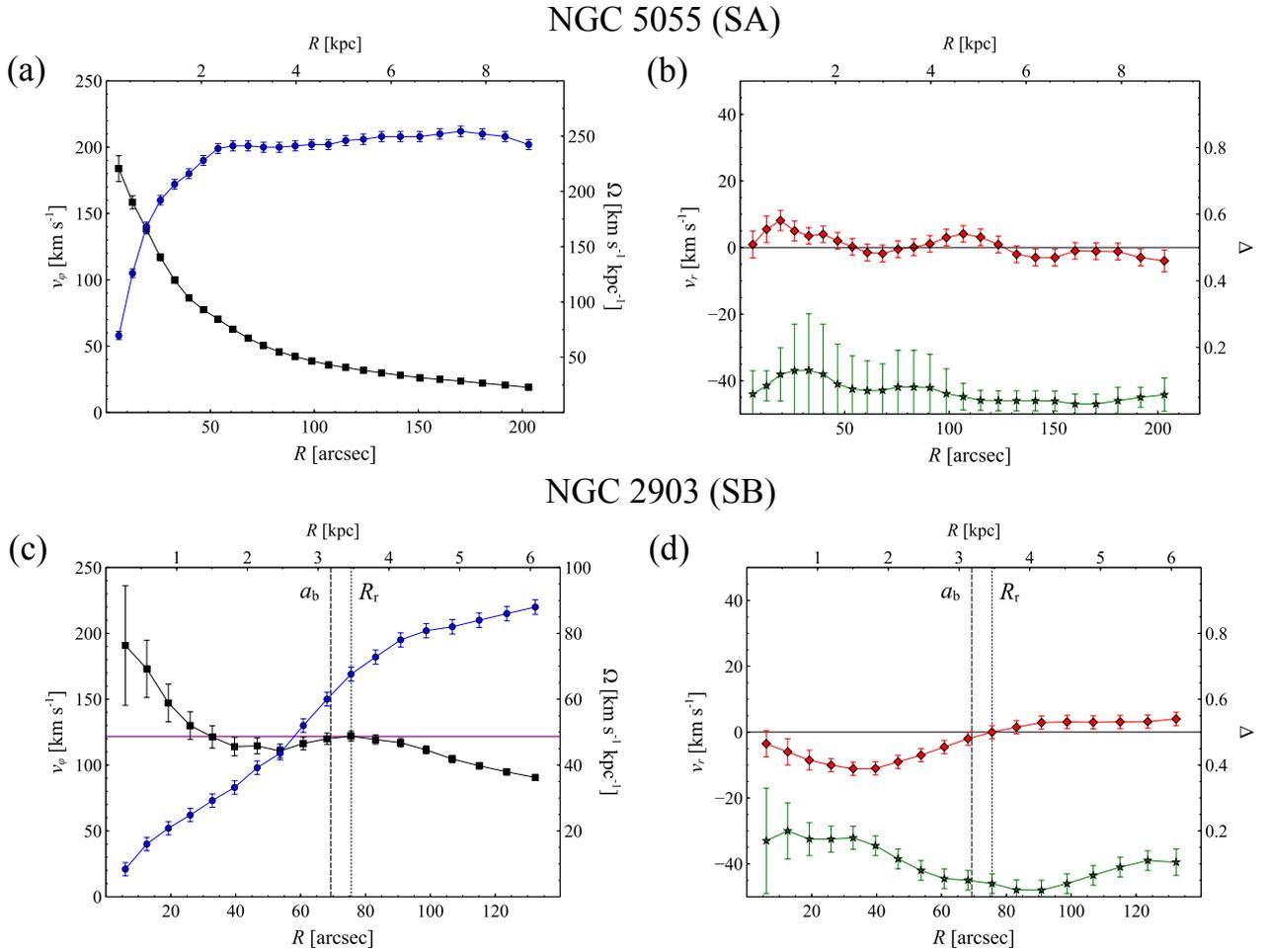} 
\end{center}
\caption{Examples of results for one SA and one SB galaxy. Panels (a) and (c) show circular velocity \(v_\varphi\) (blue circles) and angular velocity \(\Omega\) (black squares). Panels (b) and (d) show radial velocity \(v_r\) (red diamonds) and noncircular to circular velocity ratio \(\Delta\) (green stars). In panel (c), \(a_\mathrm{b}\) and \(R_\mathrm{r}\) are the bar radius and reversal radius, respectively, and the purple horizontal line in panel (c) marks the angular velocity at \(R=R_\mathrm{r}\). The reversal radius marks the location where \(v_r\) changes sign (section \ref{D11}).}
\label{fig:ex}
\end{figure}

\begin{figure}
\begin{center}
\includegraphics[width=17cm]{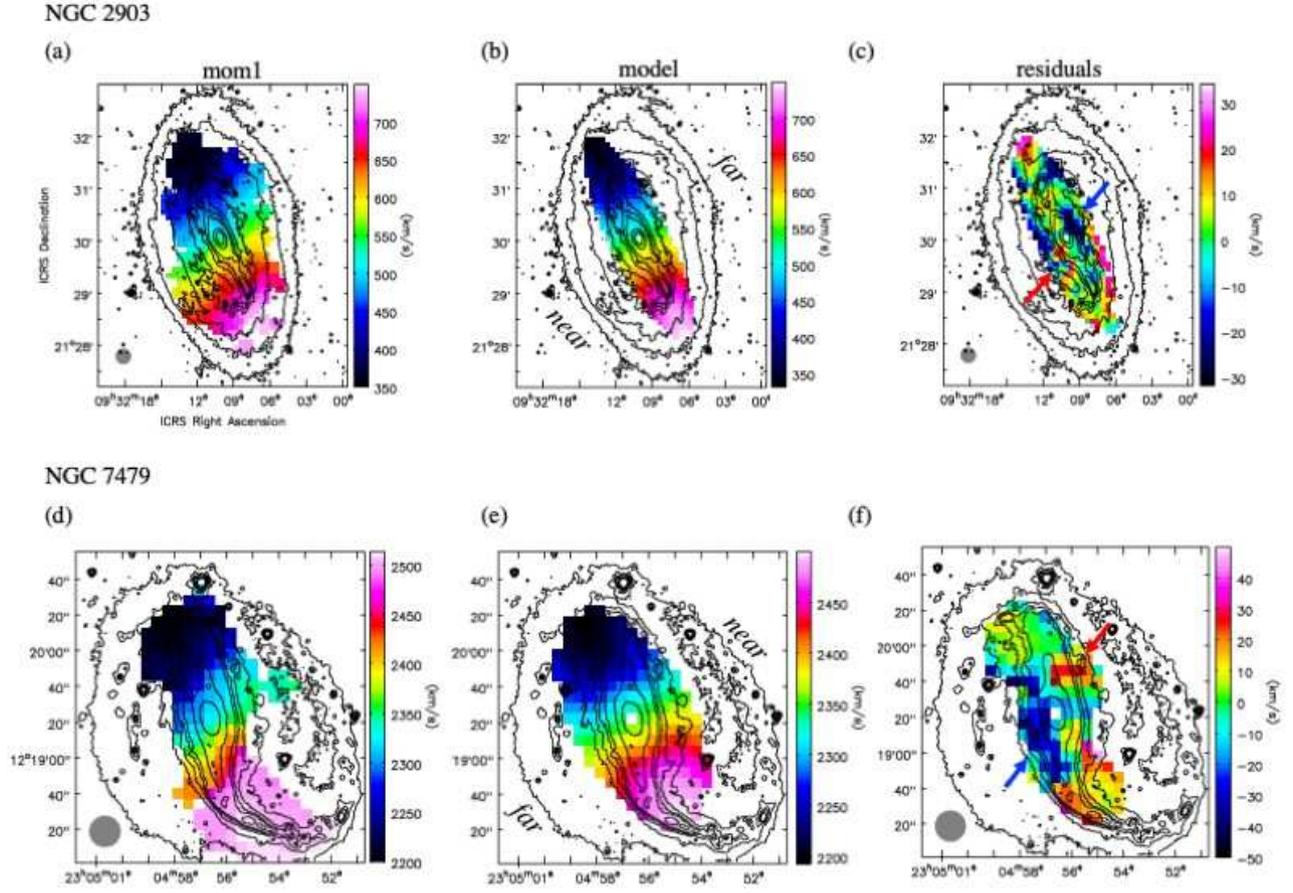} 
\end{center}
\caption{(a) Moment 1 (mean velocity) image \(\langle v\rangle\) of NGC 2903, (b) rotation model \(v_\mathrm{mod}\), and (c) residuals calculated as \(v_\mathrm{res}=\langle v\rangle -v_\mathrm{mod}\). Panels (d-f): same but for NGC 7479. The arrows in panels (c) and (f) indicate noncircular motions in the bar region. The contours are \(3.6~\micron\) \emph{Spitzer} maps acquired from NASA/IPAC Extragalactic Database. The beam size (\(17\arcsec\)) is shown at the bottom left corner.}
\label{fig:res}
\end{figure}

Having calculated the circular velocity, position angle, and inclination of the disk, it is possible to generate rotation models from the \(c_1\) term. After subtracting the models from the moment 1 images, we can obtain residuals that reveal the location of noncircular motions in the disk on kpc scale. Some examples of such investigation are shown in figure \ref{fig:res}. The residual velocity images of two SB galaxies NGC 2903 and NGC 7479 in panels (c) and (f), respectively exhibit amplitudes of the order \(|v|\sim20~\mathrm{km~s}^{-1}\) in the bar region NW and SE of the galactic center (indicated by arrows). Assuming that the spiral arms are trailing in both galaxies, as appears to be the case in the majority of barred spiral galaxies (e.g., \cite{Ath09}), it is possible from the velocity field to determine which side of the disk is closer to the observer, and distinguish inward from outward velocities in the disk. Since the near and far side of the disk are opposite in NGC 2903 and NGC 7479, the radial velocity corrected for the orientation of the disk is \(v_r<0\) in the central \(1\arcmin\) of both galaxies (see figure \ref{fig:mom1}). In section \ref{C4}, we will see that not all galaxies exhibit negative \(v_r\).

\subsection{Analysis of barred galaxies}\label{C3}

In order to analyze the molecular gas kinematics in SAB and SB galaxies, it is important to estimate the bar radius (= semimajor axis) and its position angle with respect to the position angle of the host galaxy. The orientation of bars with respect to galactic disks is random, so the apparent length of a bar can vary depending on this angle difference and inclination. In this work, the projected bar radius \(r_\mathrm{b}\), defined as the distance from the galactic center to the bar end (as in figure \ref{fig:ell}) and its position angle \(\alpha_\mathrm{b}\) were adopted from \citet{HE15}, who analyzed the \(\mathrm{S^4G}\) \emph{Spitzer} \(3.6~\micron\) data by visual inspection and searching for the radius of ellipticity maximum. The differences between the two values are usually of the order of a few percents, and we adopt a conservative uncertainty of \(10\%\) in \(a_\mathrm{b}\). The projected radius \(r_\mathrm{b}\) was converted to a deprojected bar radius \(a_\mathrm{b}\) by correcting for inclination and position angle using the equation \(a_\mathrm{b}=r_\mathrm{b}q^{-1}\sqrt{q^2\cos^2{\theta}+\sin^2{\theta}}\), where \(q\) is the ellipse flattening (see section \ref{B2}), and \(\theta\equiv\alpha_\mathrm{b}-\alpha_0\) is the difference between the position angle of the bar and that of the galactic disk.\footnote{The formula is obtained from the equation of an ellipse, \(x^2/a^2+y^2/(qa)^2=1\), expressed in polar form, where \(a=a_\mathrm{b}\), \(x=r_\mathrm{b}\cos\theta\), and \(y=r_\mathrm{b}\sin\theta\); see figure \ref{fig:ell}.} The position angles and deprojected bar radii are listed in table \ref{tab:gal1}. Note that some recent works \citep{BCO07,But15,HE15} classify some galaxies listed in table \ref{tab:gal} as SA, in contrast to SAB designation in \citet{deV91}. Since there was no reliable data for the bars in these systems, they were excluded from the sample of barred galaxies, and the final list consists of 3 SAB (NGC 3627, NGC 4303, NGC 5248) and 4 SB (NGC 613, NGC 2903, NGC 4579, NGC 7479) objects. For comparison, we include more recent classifications of galaxies in table \ref{tab:gal1}.

\begin{table}
  \tbl{Bar parameters}{
  \begin{tabular}{llcccccccc}
      \hline
      Galaxy & Morphology & \(r_\mathrm{b}\) & \(\alpha_\mathrm{b}\) & \(\theta\) & \(a_\mathrm{b}\) & \(a_\mathrm{b}\) & \(R_\mathrm{r}\) & \(\mathcal{R}\) & \(\Omega_\mathrm{b}\)  \\
      & & (arcsec) & (\degree) & (\degree) & (arcsec) & (kpc) & (arcsec) & & \((\mathrm{km~s^{-1}~kpc^{-1}})\) \\ 
     NGC & (1) & (1) & (1) & (2) & (3) & (3) & (4) & (5) & (6) \\
      \hline
      157 & SA(s)b\underline{c} & -- & -- & -- & -- & -- & -- & -- & -- \\
      613 & SB(\underline{r}s)b & \(77.1\) & \(35\) & \(8.0\) & \(80.8\) & \(10.3\pm1.0\) & \(61.0\pm6.0\) & \(0.75\pm0.11\) & \(21.5\pm2.2\) \\
      2903 & (R')SB(rs)b & \(68.1\) & \(-62\) & \(2.9\) & \(69.3\) & \(3.18\pm0.32\) & \(75.6\pm6.0\) & \(1.09\pm0.14\) & \(48.7\pm4.2\) \\
      2967 & (R')SAB(rs)c & \(8.1\) & -- & -- & -- & -- & -- & -- & -- \\
      3147 & S\underline{A}B(r'l)b & \(14.9\) & -- & -- & -- & -- & \(49.1\pm6.0\) & -- & -- \\
      3627 & SA\underline{B}(s)b pec & \(66.4\) & \(70\) & \(-6.4\) & \(67.4\) & \(2.95\pm0.29\) & \(104.0\pm6.0\) & \(1.54\pm0.18\) & \(44.3\pm3.0\) \\
      3893 & SA(s)c & -- & -- & -- & -- & -- & \(30.6\pm6.0\) & -- & -- \\
      4303 & SAB(rs)b\underline{c} & \(36.1\) & \(88\) & \(42.7\) & \(37.1\) & \(2.97\pm0.30\) & \(45.1\pm6.0\) & \(1.22\pm0.20\) & \(52\pm15\) \\
      & & & & & & & \((60.8\pm6.0)\) & \((1.64\pm0.23)\) & \((40\pm11)\) \\
      4579 & (\underline{R}L,R')SB(rs)a & \(40.7\) & \(-37\) & \(-37.3\) & \(45.2\) & \(3.62\pm0.36\) & \(61.9\pm6.0\) & \(1.37\pm0.19\) & \(52.7\pm6.6\) \\
      & & & & & & & \((76.5\pm6.0)\) & \((1.69\pm0.22)\) & \((43.9\pm5.0)\) \\
      5248 & (R')SAB(s)bc & \(27.4\) & \(38\) & \(25.3\) & \(28.2\) & \(1.78\pm0.18\) & \(45.5\pm6.0\) & \(1.61\pm0.27\) & \(73\pm16\) \\
      5678 & (R'L)SA(\underline{r}s)b pec & -- & -- & -- & -- & -- & \(25.4\pm6.0\) & -- & -- \\
      7479 & (R')SB(r\underline{s})b & \(49.3\) & \(-82\) & \(-19.3\) & \(57.0\) & \(10.2\pm1.0\) & \(>68.1\) & \(>1.09\) & \(\lesssim14\) \\
      \hline
\end{tabular}}\label{tab:gal1}
\begin{tabnote}
(1) Adopted from \citet{HE15}. The bar position angle (\(\alpha_\mathrm{b}\)) is measured anti-clockwise from west. (2) \(\theta\equiv\alpha_\mathrm{b}-\alpha_0\). (3) Deprojected bar semimajor axis. The uncertainty of \(a_\mathrm{b}\) is adopted to be \(10\%\). (4) reversal radius where \(v_r\) changes sign (secondary reversal radius is given in brackets where applicable). (5) \(\mathcal{R}\equiv R_\mathrm{r}/a_\mathrm{b}\). (6) \(\Omega_\mathrm{b}\equiv v_\varphi(R_\mathrm{r})/R_\mathrm{r}\).
\end{tabnote}
\end{table}

\begin{figure}
\begin{center}
\includegraphics[width=17cm]{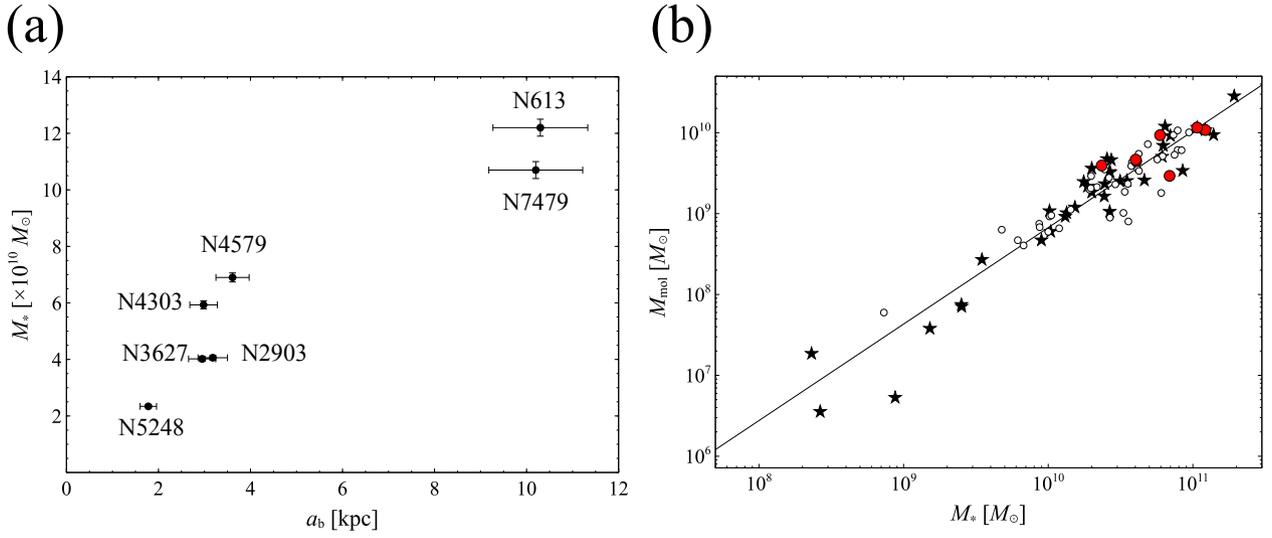}
\end{center}
\caption{(a) Total stellar mass \(M_\ast\) plotted against the bar radius \(a_\mathrm{b}\); distance uncertainties are not included in the uncertainties of \(a_\mathrm{b}\). (b) Total molecular gas mass \(M_\mathrm{mol}\) and \(M_\ast\) for all 80 barred galaxies in the COMING sample. Open circles are SAB and filled stars are SB galaxies based on the morphological classification in \citet{deV91}; the barred galaxies studied in this paper are shown as red circles (table \ref{tab:gal2}). The error bars are typically the size of the data markers. The full line shows a power law fit for all objects.}
\label{fig:bars-Ms}
\end{figure}

Figure \ref{fig:bars-Ms} shows the relations between the deprojected bar radius \(a_\mathrm{b}\), total molecular gas mass \(M_\mathrm{mol}\), and total stellar mass \(M_\ast\). Here, the molecular gas mass was calculated as \(M_\mathrm{mol}=1.36m_\mathrm{H_2}AX_\mathrm{CO}\sum_i(I_\mathrm{CO})_i\), where \(m_\mathrm{H_2}\) is the mass of a single hydrogen molecule, \(A\) is the area of a map pixel, \(I_\mathrm{CO}\equiv\int T_\mathrm{mb}dV\) is the integrated intensity, \(X_\mathrm{CO}=2\times10^{20}~\mathrm{cm^{-2}(K~km~s^{-1})^{-1}}\) is a standard conversion factor \citep{Bol13}, and summation is performed over all pixels. The factor \(1.36\) is the correction for the abundance of helium. The uncertainty of \(I_\mathrm{CO}\) was evaluated for each pixel as \(\Delta I_\mathrm{CO}=\Delta T_\mathrm{mb}\sqrt{\Delta v_\mathrm{int}\Delta v_\mathrm{ch}}\), where \(\Delta T_\mathrm{mb}\) is the r.m.s. of the emission-free channels, \(\Delta v_\mathrm{int}\) is the velocity range of the signal, and \(\Delta v_\mathrm{ch}=10~\mathrm{km~s^{-1}}\) is the channel width. Since \(M_\mathrm{mol}\propto I_\mathrm{CO}\), the uncertainty \(\Delta I_\mathrm{CO}\) was translated into the uncertainty of \(M_\mathrm{mol}\pm\Delta M_\mathrm{mol}\) \citep{Sor19}.

The total stellar mass \(M_\ast\) was calculated from the Wide-field Infrared Survey Explorer (WISE) \(3.4~\micron\) data products by \citet{Sor19}, who applied the light-to-mass conversion method described in \citet{Wen13}. The masses of the barred galaxies are listed in table \ref{tab:gal2}; the ratio \(M_\mathrm{mol}/M_\ast\) is typically of the order of \(10\%\) in the studied objects. The plots indicate that the galaxies with massive stellar disks in our sample generally tend to have longer bars and larger total molecular gas reservoirs. The relation between \(M_\ast\) and \(a_\mathrm{b}\) also holds when a larger sample is studied, as shown in \citet{DG16}. Note that the plot in figure \ref{fig:bars-Ms}(b) includes all 44 SAB and 36 SB galaxies (without SA) in the COMING sample. The relation in panel (b) can be fitted with a power law

\begin{equation}
\log{M_\mathrm{mol}}=(-3.10\pm0.48)+(1.193\pm0.046)\log{M_\ast}.
\end{equation}
Figure \ref{fig:bars-Ms}(b) shows that the 7 barred galaxies analyzed in this work are among the luminous ones (\(M_\ast>2\times10^{10}~M_\odot\)) compared to the entire sample.

\begin{table}
  \tbl{Stellar mass and molecular gas mass in the selected barred galaxies}{
  \begin{tabular}{lccc}
      \hline
      Galaxy & \(M_\mathrm{mol}\) & \(M_\ast\) & \(M_\mathrm{mol}/M_\ast\) \\
      NGC & \((M_\odot)\) & \((M_\odot)\) \\
      \hline
      613 & \((1.075\pm0.009)\times10^{10}\) & \((1.22\pm0.03)\times10^{11}\) & \(0.09\) \\
      2903 & \((4.46\pm0.02)\times10^9\) & \((4.06\pm0.09)\times10^{10}\) & \(0.11\) \\
      3627 & \((4.65\pm0.02)\times10^9\) & \((4.02\pm0.09)\times10^{10}\) & \(0.12\) \\
      4303 & \((9.33\pm0.06)\times10^9\) & \((5.93\pm0.14)\times10^{10}\) & \(0.16\) \\
      4579 & \((2.93\pm0.03)\times10^9\) & \((6.90\pm0.16)\times10^{10}\) & \(0.04\) \\
      5248 & \((3.93\pm0.02)\times10^9\) & \((2.34\pm0.05)\times10^{10}\) & \(0.17\) \\
      7479 & \((1.157\pm0.010)\times10^{10}\) & \((1.07\pm0.03)\times10^{11}\) & \(0.11\) \\
      \hline
\end{tabular}}\label{tab:gal2}
\begin{tabnote}
The molecular gas masses were derived from CO (1-0) integrated intensity, using a single conversion factor of \(X_\mathrm{CO}=2\times10^{20}~\mathrm{cm^{-2}(K~km~s^{-1})^{-1}}\) and correcting for the abundance of He. The total stellar mass was calculated from WISE \(3.4~\micron\) images \citep{Sor19}.
\end{tabnote}
\end{table}

\subsection{Gas streaming motions}\label{C4}

\begin{figure}
\begin{center}
\includegraphics[width=16cm]{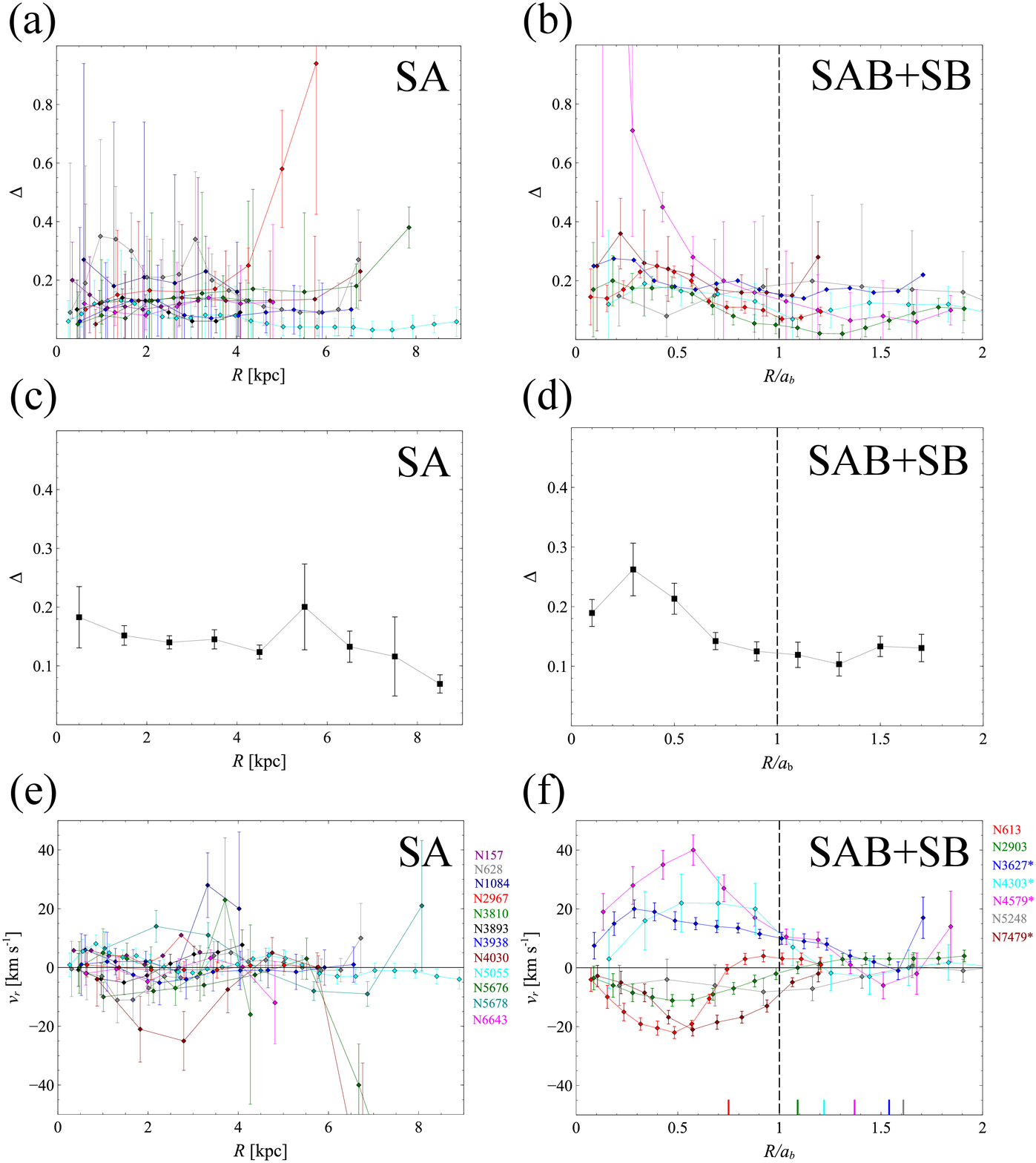} 
\end{center}
\caption{Upper panels: noncircular velocity parameter \(\Delta=v_\mathrm{nc}/c_1\) is shown for (a) non-barred and (b) barred galaxies. Middle panels: Average values of the plots in panels (a) and (b) with \(1\sigma\) uncertainty. The data were smoothed in bins of \(\delta R=1~\mathrm{kpc}\) in panel (c) and \(\delta (R/{a_\mathrm{b}})=0.2\) in panel (d). The innermost data point of NGC 4579, which has \(\Delta=2.9\), is regarded as an outlier and excluded from the smoothed curve in panel (d) (see text). Lower panels: radial velocity \(v_r\) is plotted for (e) non-barred and (f) barred galaxies. The horizontal axis in panels (b), (d), and (f) is normalized as the galactocentric radius divided by the bar radius. The color tick marks in panel (f) indicate the locations of \(R_\mathrm{r}\), and the star (e.g., N7479*) indicates objects with \(-90\arcdeg<i<0\arcdeg\) for which \(v_r\) has been corrected.}
\label{fig:noncirc}
\end{figure}

\begin{figure}
\begin{center}
\includegraphics[width=16cm]{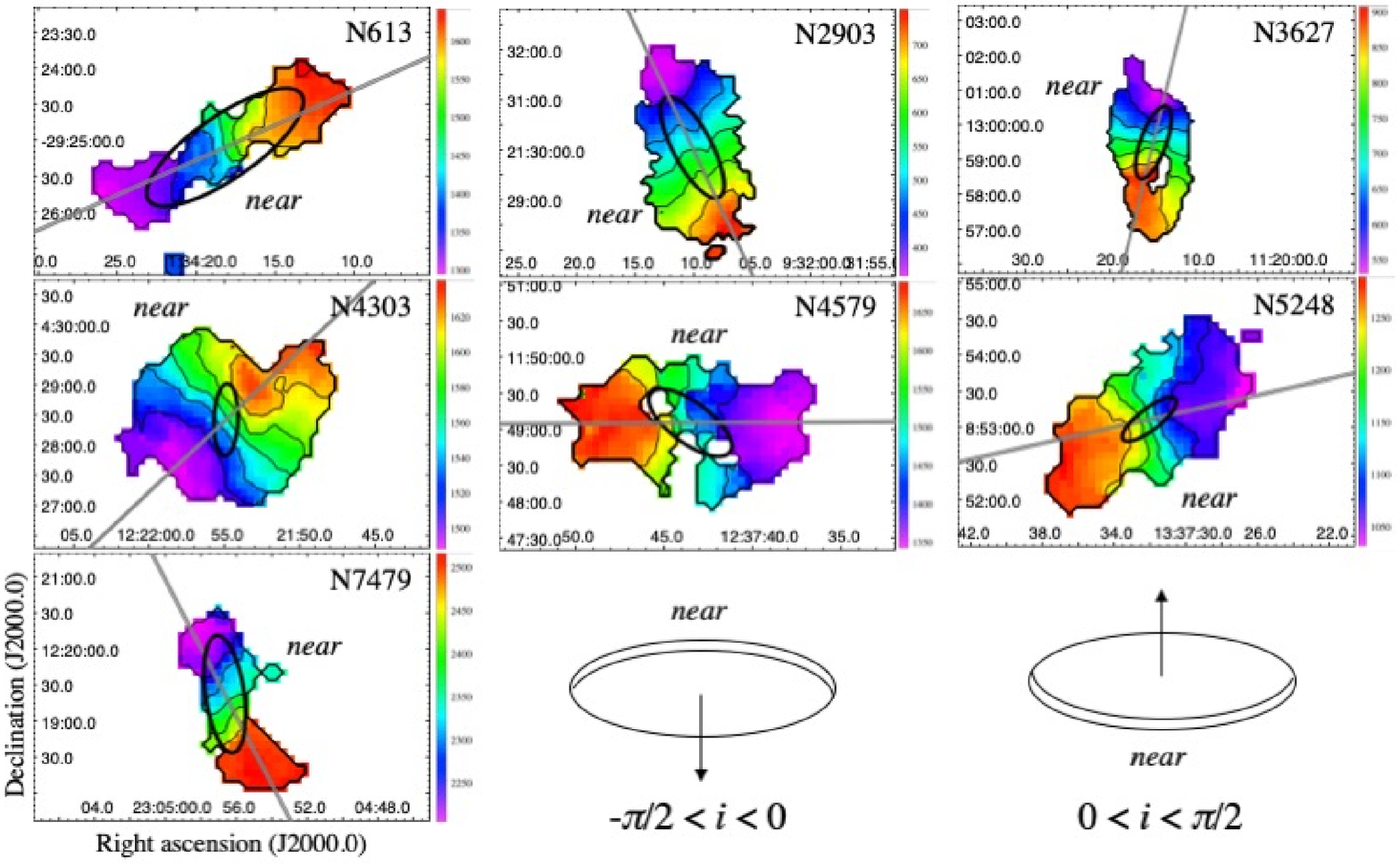}
\end{center}
\caption{Moment 1 (mean velocity in \(\mathrm{km~s^{-1}}\)) images of the barred galaxies. Shown are the calculated kinematic position angle (grey line), bar orientation (black ellipse; semimajor axis \(=r_\mathrm{b}\)), and the adopted near side of the disk with respect to the observer, assuming that the spiral arms are trailing. The near/far orientation of the disk was used to correct the sign of \(v_r=s_1/\sin{i}\) in figure \ref{fig:noncirc}(d). When the near side is at \(-180\arcdeg<\varphi<0\arcdeg\), where \(\varphi\) is measured anti-clockwise from the galactic major axis (west side), then \(0\arcdeg<i<90\arcdeg\). Five contours are plotted on each image, with equal separations from minimum to maximum values. The illustration in the bottom right shows two possible orientations of the disk from the perspective of an observer.}
\label{fig:mom1}
\end{figure}

One of the main results of the mode analysis of the velocity fields is presented in figure \ref{fig:noncirc}, where the noncircular velocity parameter \(\Delta= v_\mathrm{nc}/c_1\) and the deprojected radial velocity \(v_r=s_1/\sin{i}\) of 12 SA galaxies (on the left) and 7 SAB+SB galaxies (on the right) are plotted against galactocentric radius. The radial velocity is corrected for inclination and for the near/far side of the disk of the barred galaxies, assuming that these galaxies all have trailing spiral arms. The sign of \(i\) depends on disk orientation, such that \(0\arcdeg<i<90\arcdeg\) when the near side is at \(-180\arcdeg<\varphi<0\arcdeg\), and \(-90\arcdeg<i<0\arcdeg\) when the near side is at \(0\arcdeg<\varphi<180\arcdeg\) (figure \ref{fig:mom1}). Radial velocities uncorrected for disk orientation, i.e., as derived by the program, are shown in figures \ref{fig:plots1}-\ref{fig:plots5} in Appendix. For SA galaxies, we plot the results in their uncorrected form in figure \ref{fig:noncirc} to avoid uncertainty in the choice of orientation of their disks.

Figure \ref{fig:noncirc} shows a clear difference in the \(\Delta\) parameter between the barred and non-barred systems at 1 kpc resolution: inspecting the regions within the bar radius \(a_\mathrm{b}\) in SAB and SB galaxies, we find a maximum of \(\Delta\sim0.26\) at \(R/a_\mathrm{b}\sim0.3\), which is a factor of 1.5-2 larger than in SA systems, where no significant variation with radius is observed. On the other hand, the values appear to be similar (\(\Delta\sim0.10\mathrm{-}0.15\)) in both barred and non-barred objects at large radii.

The radial velocities \(v_r\), plotted in figure \ref{fig:noncirc}(f), indicate that the rise in \(\Delta\) can be attributed to some extent to the \(s_1\) coefficient in \(v_\mathrm{nc}\). Molecular gas is affected by the presence of the bar and exhibits noncircular (streaming) motions with radial velocity amplitudes typically \(|v_r|\sim20~\mathrm{km~s}^{-1}\) at 1 kpc resolution. Note that both \(v_r<0\) and \(v_r>0\) velocities are observed. The reason for this is that there can be both inward and outward motions of gas in the bar region as there are negative and positive gravitational torques (tangential forces) in the vicinity of the bar that give rise to tangential acceleration \(|\ddot{\varphi}|>0\) and \(|v_r|>0\); gas is accelerated as it approaches the bar on the upstream side and decelerated on the downstream side. The streaming motions are clearly visible in figure \ref{fig:mom1}, where isovelocity contours tilt with respect to kinematic minor axis in the same sense as the bar. If the gas orbits were ellipses, the isovelocities in the bar would remain without tilt only if \(\theta=0\) or \(\theta=\pi/2\) \citep{Kal78}. However, we find substantially tilted isovelocity contours even in NGC 2903, where \(\theta\approx0\). This behavior can be explained in terms of gas inflow in the bar region.

Among the barred galaxies, NGC 4579 exhibits notably peculiar CO (1-0) velocities with a large value of \(\Delta\) in the innermost ring. The high fraction of noncircular motions appears to be related to a blueshifted velocity component in the galactic center, as evident in the moment 1 image in figure \ref{fig:mom1}. Similar peculiar motions of gas can also be seen in the velocity field of ionized gas traced by [O III] in the central \(R<20\arcsec\) \citep{Dum07}, as well as within \(R<30\arcsec\) of the interferometric CO (1-0) data in \citet{Sof03}.

NGC 3147 is not included in the plots because it is very weakly barred (classified as \(\mathrm{S\underline{A}B}\) or SA). The bar length in the catalogue of \citet{HE15} suggests a nuclear bar (\(r_\mathrm{b}\sim14.9\arcsec\)); similarly, \citet{Cas08} reported a radius of \(r_\mathrm{b}\sim7.5\arcsec\), which is too small to be resolved in our images. Nevertheless, the result of Fourier decomposition of this galaxy (Appendix) exhibits noncircular motions that peak at the innermost radii, in agreement with other barred systems.

Similarly, NGC 5678 is classified as SAB in \citet{deV91} and as SA in \citet{HE15}. There is no clear indication of a bar in this galaxy, so we place it in the SA plot in figure \ref{fig:noncirc}(e). However, we note that the behavior of \(v_r\) exhibits a change in sign at a radius of \(R\sim4~\mathrm{kpc}\), indicating that the velocity field is perturbed. The optical/near-infrared morphology suggests asymmetric morphology with an outer pseudoring.

On the other hand, although NGC 2967 is classified as SA in \citet{deV91}, ellipticity measurements reported in \citet{HE15} and \citet{Sal15} indicate an SAB type; the bar radius, however, is only \(\sim8\arcsec\), which is smaller than the resolution of our images and the influence of this nuclear bar can be neglected. This is supported by the fact that \(v_r\) behaves as expected from an SA (figure \ref{fig:plots2}).

\section{Discussion}\label{D}

\subsection{Corotation resonance and bar pattern speed}\label{D1}

Based on analytical models of weak bar potentials,\footnote{A commonly used definition of a weak bar is the condition \(|\Phi_\mathrm{b}/\Phi_0|\ll1\), where \(\Phi_0\) is the axisymmetric component and \(\Phi_\mathrm{b}\) is the non-axisymmetric component of the potential.} a bar is expected to extend between the radius of an inner Lindblad resonance (if there is any) and the corotation radius \(R_\mathrm{CR}\), where the disk material and bar rotate with the same angular velocity (\(\Omega_{R=R_\mathrm{CR}}=\Omega_\mathrm{b}\)) \citep{Con80,Elm96,BT08}. It is therefore expected that the corotation radius in the majority of weakly barred galaxies is at least as large as the bar radius. Numerical simulations also suggest that the ratio of the corotation radius to the bar radius, defined as the parameter \(\mathcal{R}\equiv R_\mathrm{CR}/a_\mathrm{b}\), is typically in the range \(\mathcal{R}=1.0\mathrm{-}1.4\) \citep{Ath92}; these are referred to as the fast bars, in contrast to the slow bars, which are loosely defined as having \(\mathcal{R}\gtrsim1.4\). On the other hand, since the gravitational torques generated by the bar are generally smaller at radii \(R>R_\mathrm{CR}\) compared to the bar region \citep{DG16}, the velocity dispersion of gas is expected to decrease at the corotation radius. In accordance with the density wave theory, the velocity field may even reverse phase at \(R\sim R_\mathrm{CR}\), in the sense that radial streaming motions of gas associated with spiral arms change sign at resonance radii, such as corotation and outer Lindblad resonance (OLR), in systems where the spiral pattern is forced by a bar \citep{Kal78}. Therefore, identifying the radii where the velocity reversals occur can yield the locations of resonances.

\subsubsection{Radial velocity reversal}\label{D11}

The argument of \citet{Kal78} was recently applied by \citet{Fon11} on high-resolution H\(\alpha\) data to develop a new technique of estimating the pattern speed (resonance radii) in spiral galaxies. In \citet{Fon17}, \(\Omega_\mathrm{b}\) is found indirectly by first measuring \(R_\mathrm{CR}\), and since the angular velocity of the disk is equal to that of the bar at \(R_\mathrm{CR}\), the value of \(\Omega\) at that radius becomes the bar pattern speed. Here, \(R_\mathrm{CR}\) is determined at the radial range where the velocity residuals, obtained by subtracting a model velocity field image from the H\(\alpha\) velocity field image, are minimal and reversals are identified. In principle, the method can also be applied to estimate the pattern speed of spiral arms if they are regarded as density waves \citep{Fon14}.

The possibility that the gas velocity field images (e.g., CO, HI, H\(\alpha\)) can be used to determine the corotation radius \(R_\mathrm{CR}\) of spiral arms, based on the assumption that they are density waves, was also discussed in \citet{Can93}, \citet{GB94}, \citet{CA97}, \citet{Col14}, and others. For example, high-resolution (\(\sim1\arcsec\)) images of the grand-design spiral galaxy M51, analyzed by \citet{Col14} and \citet{Que16}, reveal streaming motions in the spiral arms that can be associated with different wave modes and corotation, although possibly amplified by tidal interaction with M51b.

Compared to spiral arms, the bar instability is believed to be a genuine, long-lasting density wave in the disk and the response of gas in terms of streaming motions is more obvious, so that the location of its corotation can be estimated by applying the following simplified method. Since gas streaming motions are present in the bar region, we will observe \(|v_r|>0\) inside \(R_\mathrm{CR}\) and \(v_r\approx0\) near it, and the spatial extent of the \(m=1\) mode is determined by the reversal radius \(R_\mathrm{r}\) in the axisymmetric radial velocity \(v_r\) (table \ref{tab:gal1}). The existence of \(v_r\approx0\) regions near the corotation can be understood from the simulations of \citet{Ath92}. For example, figure 2(b) of her paper shows that velocity vectors have magnitudes close to zero in the corotation region in the reference frame where the bar is at rest. Since the material here then must rotate about the galactic center at the same rate as the bar, the rotation is nearly circular and both radial velocity \(v_r\) projected along the minor galactic axis and radial velocity dispersion \(\sigma_{v_r}\) are expected to be minimal. Note, however, that these models consider isolated (non-interacting) barred galaxies.

In order to examine the velocity reversal method for barred galaxies, we first discuss on the behavior of \(v_r\) determined by mode analysis. Although OLR also is an important resonance, often related to the formation of outer rings in galaxies \citep{BC96,BCO07}, it may lie outside the region where CO (1-0) is firmly detected in our images, so in this section we discuss corotation only. The location of OLR and its relation with the observables is briefly addressed in section \ref{D3}.

As evident in figures \ref{fig:ex} and \ref{fig:noncirc}, the calculated \(v_r\) exhibits a reversal at a specific radius \(R_\mathrm{r}\) in the case of barred galaxies, in contrast to the majority of non-barred spiral galaxies, where \(v_r\approx0\) at most radii. In all but one galaxy, the reversal radius \(R_\mathrm{r}\) is somewhat larger than the bar radius (table \ref{tab:gal1}), which is comparable to the expected value of \(R_\mathrm{CR}\) based on analytical analyses and simulations of weakly barred galaxies. Assuming that \(R_\mathrm{r}\sim R_\mathrm{CR}\), we get \(\overline\mathcal{R}=1.24\pm0.12\) (\(1\sigma\) of the mean), with a wide range \(\mathcal{R}=0.8\mathrm{-}1.6\), in general agreement with previous measurements that apply different methods (e.g., \cite{ADC03,RSL08,Cor11,Agu15,Fon17}). For example, the simulations in \citet{RSL08} yield a similar value of \(\overline\mathcal{R}=1.44\pm0.29\) for intermediate (SBb) barred galaxies, comparable to the ones in our sample. Although the uncertainty is large, arising mainly from the uncertainty of \(R_\mathrm{r}\), the values suggest the presence of both fast and slow rotator bars in intermediate barred spirals. We also note that the minimum of the smoothed \(\Delta\) in SAB and SB systems in figure \ref{fig:noncirc}(d) lies at \(R/a_\mathrm{b}\sim1.3\).

The corotation of a trailing spiral pattern is expected to be marked by a reversal from negative (inward) to positive (outward) \(v_r\) \citep{Kal78}. Figure \ref{fig:noncirc}(f) indicates that NGC 4303 and NGC 4579 may have two reversals, one just beyond the bar end and the other farther out in the spiral arms. Since both galaxies exhibit \(v_r<0\) between the first and second reversal radii, this region may be inside the corotation of the spiral pattern. For instance, similar gas dynamics with streaming motions and reversals in the spiral arms is observed in the SB galaxy NGC 1365 \citep{LLA96,EGA09}.

In the discussion that follows, we associate the \(v_r\)-reversal radius \(R_\mathrm{r}\) (where \(v_r=0\); radial velocity decreases to zero or changes sign) with \(R_\mathrm{CR}\) as an approximate location of the corotation resonance, adopting an uncertainty given by the resolution of the image grid (\(\Delta R_\mathrm{r}=6\arcsec\)). The error of \(q\), discussed in section \ref{C1} (figure \ref{fig:pa-incl}), is the dominant contributor to the uncertainty of the deprojected radius, and this is reflected on \(R_\mathrm{r}\). In galaxies with large \(\theta\), the projection effects nearly cancel out for \(R/a_\mathrm{b}\), since both \(R\propto q^{-1}\) and \(a_\mathrm{b}\propto q^{-1}\), but can be large in objects with small \(\theta\). In the extreme case of \(\theta=0\arcdeg\), the estimated bar radius is \(a_\mathrm{b}=r_\mathrm{b}\), but we still have \(R_\mathrm{r}\propto q^{-1}\) and the largest errors arise for small-\(q\) (close to edge-on) systems. If we take NGC 2903 as an example, figure \ref{fig:pa-incl} (see also table \ref{tab:gal1}) shows that the photometric \(q\) is larger than the kinematic one. If we Fourier decomposed this velocity field by using the photometric \(q\), we would get somewhat smaller \(R_\mathrm{r}\). From table \ref{tab:gal1}, the uncertainty of \(6\arcsec\) corresponds to \(\sim10\mathrm{-}20\%\) of the bar radius in all objects. The total uncertainty of the bar pattern speed is estimated to be of the order of \(\sim20\%\).

\subsubsection{Comparison with other studies}\label{D12}

Although our measurements of \(\Omega_\mathrm{b}\) are not the first for most of the selected galaxies, the applied \(v_r\)-reversal method is new, relatively simple to perform, and applicable to 1-kpc resolution data. In this section, we demonstrate that there are no significant discrepancies compared to previous works.

The derived value of the pattern speed in NGC 3627 is compatible with \(\Omega_\mathrm{b}=50^{+3}_{-8}~\mathrm{km~s^{-1}~kpc^{-1}}\) found by \citet{Hel03} who applied the Tremaine-Weinberg method. Similarly, the parameter \(\mathcal{R}\), which is distance-independent, is comparable to \(\mathcal{R}\sim1.6\) in \citet{Hir09}. We also find a marginal agreement for the value of \(\mathcal{R}\sim1\) for NGC 613 in \citet{EEM92}, who estimated the location of \(R_\mathrm{CR}\) from an analysis of optical morphology. The values of \(\mathcal{R}\) for both NGC 4303 and NGC 4579 are consistent with the values of \(\mathcal{R}=1.70\pm0.45\) and \(\mathcal{R}=1.46\pm0.30\), respectively, in \citet{RSL08}, who simulated model galaxies. We also find agreement with \(\mathcal{R}=1.07\pm0.09\) and \(\Omega_\mathrm{b}=49.6^{+3.2}_{-2.9}~\mathrm{km~s^{-1}~kpc^{-1}}\) \citep{Fon17}, as well as \(\Omega_\mathrm{b}=53~\mathrm{km~s^{-1}~kpc^{-1}}\) for NGC 4303 \citep{KS06}, who adopted a similar distance to the galaxy.

The low derived bar pattern speed of NGC 7479 (its upper limit) is comparable to the value of \(\Omega_\mathrm{b}\approx18~\mathrm{km~s^{-1}~kpc^{-1}}\) found by \citet{Fat09}, who applied the Tremaine-Weinberg method on H\(\alpha\) data and adopted a similar distance to the galaxy. \citet{Fon17} obtained the same value of \(\Omega_\mathrm{b}\) and \(\mathcal{R}=1.14\pm0.05\), as did \citet{SCC95}, both consistent with our lower limit.

The case of NGC 5248 is less clear because there are opposite views on the bar length in this system. From inspection of an \(R\)-band image, \citet{Jog02} claimed the presence of a much larger bar (\(a_\mathrm{b}\sim95\arcsec\)) with \(\mathcal{R}\sim1.2\), that resembles an oval disk in which a spiral pattern is embedded. However, \citet{HE15} and \citet{Sal15} find an ellipticity maximum at a much smaller radius, compatible with a small bar. The \(v_r\) derived from the CO (1-0) velocity field exhibits a reversal at \(\mathcal{R}\sim1.6\), and figures \ref{fig:noncirc}(f) and \ref{fig:mom1} suggest that the velocity field in this galaxy seems to be only mildly disturbed over most radii.

\subsubsection{Bar pattern speed of NGC 2903}\label{D13}

Among the examined barred galaxies, NGC 2903 is an SB with the position angle of the bar (\(\alpha_\mathrm{b}\)) found to be nearly equal to that of the disk (\(\alpha_0\)) of the host galaxy as determined in our analysis (table \ref{tab:gal1}). This condition allows us to investigate the bar pattern speed by using an alternative approach, as demonstrated in \citet{Sal16} for NGC 1808 [see also \citet{Hir09}].

As mentioned in section \ref{B2}, the measured mean velocity can be expressed as \(v=v_\mathrm{sys}+v_\varphi\sin{i}\cos{\varphi}+v_r\sin{i}\sin{\varphi}\). Assuming that molecular clouds in the bar region propagate in nearly radial direction (in the non-inertial reference frame of the bar) in the ``offset ridges'' where prominent dust lanes are often observed, and that the dust lanes are the density maxima in gas flow, the projected azimuthal velocity (in the inertial frame of the observer) in the bar region is much larger than the radial velocity component along the line of sight, hence the equation simplifies to \(v\approx v_\mathrm{sys}+v_\varphi\sin{i}\). Under this condition, the observed azimuthal velocity is related to the rotation of the bar via \(v_\varphi=R\Omega_\mathrm{b}\cos{(\alpha_\mathrm{b}-\alpha_0)}\approx R\Omega_\mathrm{b}\).

The expected linear relation \(v_\varphi\propto R\) can be seen clearly in the plot of \(v_\varphi\) and \(\Omega\) in figure \ref{fig:ex}(c), where the angular velocity of the disk appears nearly constant within a wide range of \(30\arcsec<R<80\arcsec\) that marks the extent of the bar. The flat portion of the angular velocity curve here can be interpreted as the pattern speed of the bar in NGC 2903, and is consistent with the value of \(\Omega_\mathrm{b}=48.7\pm4.2~\mathrm{km~s^{-1}~kpc^{-1}}\) determined from the \(v_r\) reversal method in section \ref{D11}, where we have assumed that \(R_\mathrm{r}\simeq R_\mathrm{CR}\) (table \ref{tab:gal1}). In fact, figure \ref{fig:ex}(c) shows that the angular velocity \(\Omega\) is lower than the derived value of \(\Omega_{R=R_\mathrm{r}}\) at some radii inside corotation. Since the bar is rotating slower than the disk at \(R<R_\mathrm{CR}\), the bar pattern speed is possibly somewhat lower, in better agreement with \citet{Hir09}.

\subsection{Bar radius, total stellar mass, and bar pattern speed}\label{D2}

\begin{figure}
\begin{center}
\includegraphics[width=17cm]{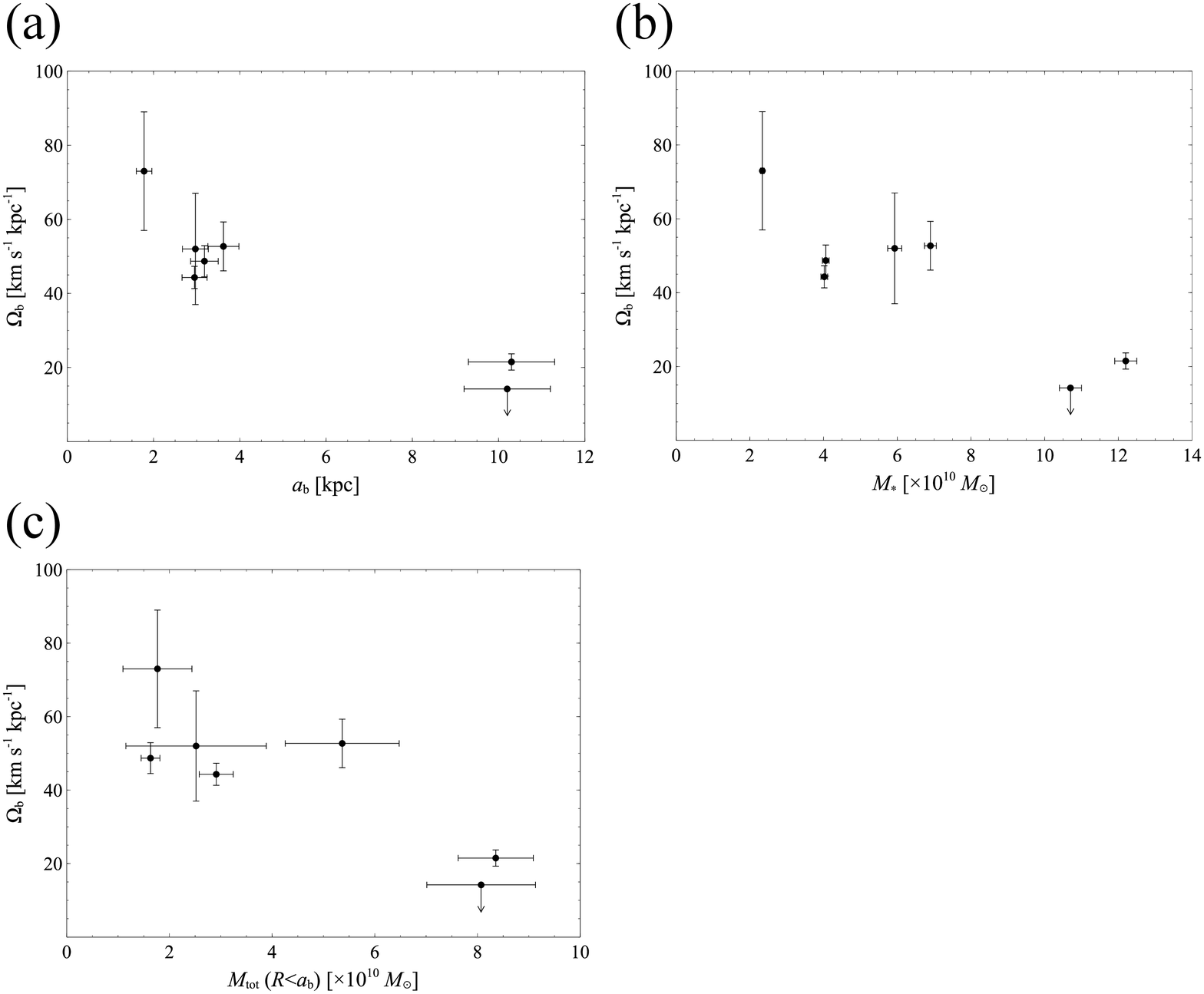} 
\end{center}
\caption{Angular velocity \(\Omega_\mathrm{b}\) at the reversal radius \(R_\mathrm{r}\) plotted against (a) bar radius \(a_\mathrm{b}\), (b) total stellar mass \(M_\ast\), and (c) dynamical mass (baryonic + dark matter) within the bar radius, \(M_\mathrm{tot}(R<a_\mathrm{b})\).}
\label{fig:bars}
\end{figure}

Using the estimated bar pattern speed \(\Omega_\mathrm{b}\), we now examine its relation with two fundamental parameters of the host galaxies: bar radius \(a_\mathrm{b}\) and total stellar mass \(M_\ast\) (figure \ref{fig:bars}). Although the sample is small, \(\Omega_\mathrm{b}\) is found to be relatively low in systems with large bars and massive stellar disks. The first result, shown in panel (a), is not entirely surprising: if the rotational velocities of the galaxies are similar and approximately flat or decreasing at large radii, then the angular velocities are decreasing with \(R\); if \(R_\mathrm{r}\) is large, as is found in systems with long bars, \(\Omega_\mathrm{b}\) is expected to be relatively low. This trend is generally expected from simulations \citep{DS00} and analytical analyses \citep{LB79}.

Figure \ref{fig:bars}(b) tentatively suggests a negative correlation between the bar pattern speed and the total stellar mass \(M_\ast\) in these systems. If large-\(M_\ast\) galaxies have massive dark matter halos, the observed trend may indicate that dynamical friction on bars by the dark matter halos effectively slows them down \citep{Cha43,TW84b}. From the point of view of numerical simulations, the following scenario is expected: as galaxies evolve, their masses increase and the rotational speeds of bars decay due to frictional torques exerted by dark matter halos and angular momentum transfer to the outer disk and bulge \citep{Sel80,Wei85,DS00,Ath03,Min12,Sah12}. In the sample, we have two SB galaxies with massive stellar disks and slowly rotating bars: NGC 613 and NGC 7479. Curiously, we find \(\mathcal{R}\sim1\) in NGC 613, which classifies it as a \emph{fast} bar. This would be expected if the galaxy hosts a ``maximum disk'' with a minimal fractional contribution of the dark matter halo in the bar region (\(R\lesssim R_\mathrm{CR}\)) \citep{DS00}. Some studies suggest that maximum disks are indeed more common in luminous spiral galaxies, such as these in our sample, compared to low-luminosity ones \citep{KSR03}.

If the baryonic mass in the studied galaxies is proportional to the dark matter mass, the negative relation in figure \ref{fig:bars}(b) would be observed for both \(M_\ast\) and the total galactic mass (baryonic + dark matter) \(M_\mathrm{tot}\). The rotation curves of the 7 barred galaxies, which can be used as proxies of the total (dynamical) mass within a certain radius (assuming symmetrical distribution of mass) via \(M_\mathrm{tot}(<R)=v^2_\varphi R/G\), attain maximal measured values in a range between \(160~\mathrm{km~s}^{-1}\lesssim v_\varphi\lesssim270~\mathrm{km~s}^{-1}\), with no clear correlation with \(M_\ast\). This is possibly because \(M_\ast\) was derived within a larger spatial extent than \(M_\mathrm{tot}\). On the other hand, we find a negative correlation between \(\Omega_\mathrm{b}\) and the dynamical mass within the bar radius, \(M_\mathrm{tot}(R<a_\mathrm{b})\), as shown in figure \ref{fig:bars}(c). To further investigate the extent of coupling between \(M_\ast\) and \(M_\mathrm{tot}\) with \(\Omega_\mathrm{b}\), it is important to study the shapes of rotation curves of these galaxies at larger radii.

\subsection{The effects of galactic mass distribution and tidal interactions on resonance radii}\label{D3}

In the simplistic discussion above, the barred galaxies are regarded as isolated and similar in terms of mass distribution and underlying rotation curves. However, recent numerical simulations give insight into possible complications in systems undergoing tidal interactions during fly-bys, as well as subtleties of gas kinematics and the location of corotation that arise from differences in rotation curves (mass models) of the host galaxies \citep{PW18}.

\subsubsection{Gas kinematics in simulated barred galaxies}

\begin{figure}
\begin{center}
\includegraphics[scale=0.6]{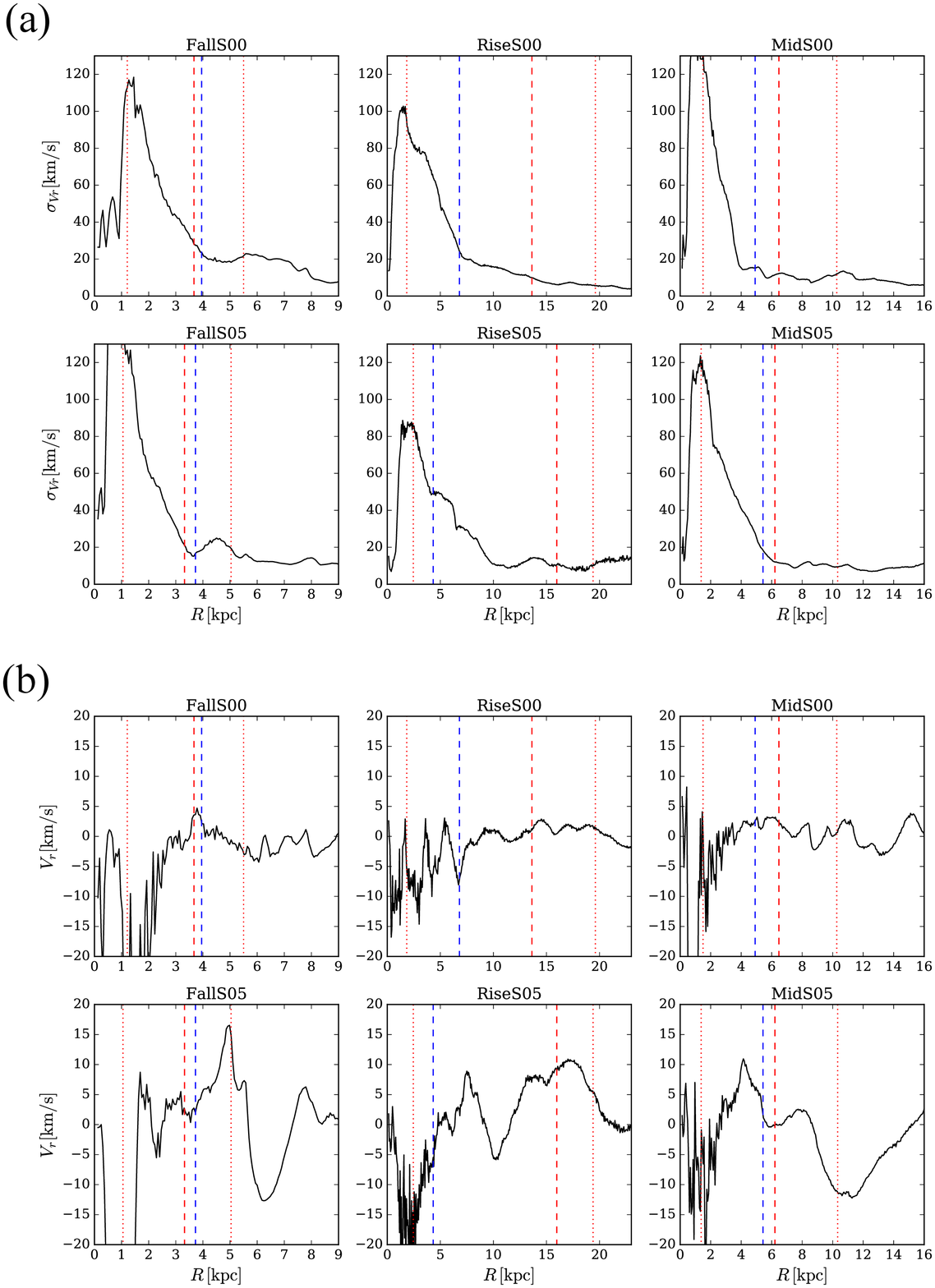}
\end{center}
\caption{(a) Azimuthally-averaged radial velocity dispersion \(\sigma_{V_r}\) of gas in isolated bars (upper panels) and weakly interacting systems (lower panels) calculated for simulated galaxies in \citet{PW18} at time \(t=2.4~\mathrm{Gyr}\). The column names refer to the rotation curve models (Fall, Rise, Mid) as defined in their paper. The bar radius \(a_\mathrm{b}\) and corotation radius \(R_\mathrm{CR}\) are indicated by vertical blue and red dashed lines, respectively. Dotted lines show the radii of Lindblad resonances. (b) Azimuthally-averaged radial velocity \(V_r\).}\label{fig:sigma_vr}
\end{figure}

In figure \ref{fig:sigma_vr}, we show an example of gas kinematics calculated for simulated galaxies in \citet{PW18} for different rotation curve models (referred to as the Fall, Rise, and Mid models in figure 1 of their paper; the distribution of gas is shown in figure 26) and interaction strength: S00 for isolated systems and S05 for weakly interacting ones. In panel (a), we show the azimuthally averaged radial velocity dispersion \(\sigma_{V_r}\) as a function of galactocentric radius, the bar semimajor axis \(a_\mathrm{b}\), and the corotation radius \(R_\mathrm{CR}\) at time \(t=2.4~\mathrm{Gyr}\).\footnote{The azimuthally averaged radial velocity in simulated galaxies is written as \(V_r\) to distinguish it from the axisymmetric radial velocity derived from Fourier decomposition, which we write as \(v_r\). The velocity dispersion is \(\sigma_{V_r}=\sqrt{\langle V_{r,\varphi}^2\rangle-\langle V_{r,\varphi}\rangle^2}\), where \(\langle\rangle\) means averaging in azimuth \(\varphi\).} Note that the bar length is determined by tracing the linearity of the bar figure such that a deviation of a Gaussian fit by \(10\arcdeg\) from the one in the inner 2 kpc is regarded as the beginning of a spiral arm/bar end at that radius \citep{PW18}. This definition is different from the one adopted here for the observed bars \citep{HE15}.

The results show that, in general, the corotation radius is often found to be in the vicinity of a local minimum in \(\sigma_{V_r}\) (and just outside the region of high \(\sigma_{V_r}\)), as discussed in section \ref{D11}, but may be as much as \(\sim30\%\) offset, especially in galaxies with rising rotation curves. Note that the region of high \(\sigma_{V_r}\) is comparable to \(\Delta\) in section \ref{C4} and figure \ref{fig:noncirc}, where it was found that an increase in the ratio of noncircular to circular motions in SAB and SB systems, quantified by \(\Delta\), is located within \(R/a_\mathrm{b}\lesssim1\). Figure \ref{fig:sigma_vr}(a) suggest that the global maximum of \(\sigma_{V_r}\) is generally found in the vicinity of the inner Lindblad resonance, at a radius of \(R/a_\mathrm{b}\sim0.3\), which is notably similar to the behavior of \(\Delta\) in figure \ref{fig:noncirc}(d). In some models (FallS00 and MidS00), another small jump is observed in the vicinity of the outer Lindblad resonance.

In figure \ref{fig:sigma_vr}(b), we show azimuthally averaged radial velocity \(V_r\) for the same galaxy models as in panel (a). The gas flow is generally negative (inward) with \(V_r\lesssim0\) within \(R<R_\mathrm{CR}\) and \(V_r\approx0\) beyond \(R>R_\mathrm{CR}\) in all isolated galaxy types, in agreement with the assumptions made in section \ref{D11}. In fact, \(R_\mathrm{CR}\) is found to be at least at the reversal radius and in some cases farther out (\(R_\mathrm{CR}\gtrsim R_\mathrm{r}\)). Although the offsets seem to be rather small (\(\sim20\%\)) especially in FallS00 and RiseS00 models, corotation does not always mark a clear reversal in \(V_r\). Moreover, note that the situation may dramatically change in systems undergoing weak interactions. In all S05 plots in figure \ref{fig:sigma_vr}, the radius \(R_\mathrm{r}\) seems to be uncorrelated with \(R_\mathrm{CR}\), though we still find that \(R_\mathrm{CR}>R_\mathrm{r}\) in all cases. The plots also show that \(V_r\) appears to attain its minimum in the vicinity of the inner Lindblad resonance in all models.

\subsubsection{Comparison with observed molecular gas kinematics}

The parameters of individual galaxies presented in section \ref{C} and Appendix can be compared with the simulations. The input information is the shape of the rotation curve and the presence or absence of a tidally interacting neighbor. In addition to relatively isolated galaxies (NGC 613, NGC 2903, NGC 5248, and NGC 7479), the sample presented here includes one member of a compact group (NGC 3627; Leo Triplet), as well as cluster members (NGC 4303 and NGC 4579; Virgo cluster); we also find a variety of rotation curve shapes, which are mainly Fall or Mid types, as judged from published HI data, such as THINGS \citep{dB08}, that trace the dynamics to larger radial extent. These two models correspond to disk-dominated systems with \(\mathcal{R}<1.4\), unlike Rise which is dark matter dominated and generates slow rotators (\(\mathcal{R}>1.4\)) in the simulations.

Comparing \(v_r\), derived from Fourier decomposition of CO velocity fields, with \(V_r\) in figure \ref{fig:sigma_vr}, we find that the models for isolated bars resemble the behavior of radial velocity in the galaxy sample, i.e., generally \(V_r\approx0\) at \(R=R_\mathrm{CR}\). As an example, we compare the results of simulations with the above discussed SB system NGC 2903, which has \(\theta\approx0\) (bar is aligned with the galactic major axis) so that we can compare \(v_r\) directly with \(V_r\). The panel in figure \ref{fig:ex}(c) suggests a rising rotation curve; however, HI measurements beyond \(R\sim7~\mathrm{kpc}\) clearly indicate a Fall type \citep{dB08}. A comparison of the measured values with the simulation results for FallS00 reveals the following similar features: (1) the corotation radius is close to the bar radius, (2) this radius is in the vicinity of a steep rise in \(\sigma_{V_r}\) and \(\Delta\), and (3) the radial velocity is \(v_r\lesssim 0\) within \(R\lesssim R_\mathrm{CR}\). Similar behavior is observed in other isolated galaxies too, such as NGC 613 and NGC 7479.

Another example is NGC 3627 (M66), a member of the Leo Triplet galaxy group that includes NGC 3623 (M65) and NGC 3628. The rotation curve of NGC 3627 can be regarded as a Mid type based on HI kinematics \citep{dB08}, and due to its interaction past \citep{ZWA93}, we may consider both MidS00 and MidS05 panels of figure \ref{fig:sigma_vr}. Compared to figure \ref{fig:plots2}, we recognize that \(R_\mathrm{r}\) lies at a radius larger than \(a_\mathrm{b}\) and that \(\sigma_{V_r}\) and \(\Delta\) are at minimal values. Note that NGC 3627 exhibits \(v_r>0\) at some radii, as evident from figure \ref{fig:noncirc}(f), even though it also has a relatively small \(\theta\). This behavior is comparable to \(V_r\) in the MidS05 model.

Similarly, figure \ref{fig:noncirc}(f) shows that NGC 4303 and NGC 4579 also exhibit some outward motion of molecular gas at \(R<R_\mathrm{r}\). This can be understood if the clouds are accelerated inside corotation due to positive torques exerted by the bar on the upstream side. Curiously, these two objects are members of the Virgo cluster, which makes them good candidates for subtle tidal interactions, as in all S05 models in simulations. For instance, NGC 4303 (M61) may be a MidS05 type, judging from the large-scale rotation curve in \citet{Sof99}, and its optical morphology (dust lanes in high resolution images) resembles the distribution of gas in figure 26 (MidS05) of \citet{PW18}. Figure \ref{fig:sigma_vr}(b) shows that weak interactions may notably elevate the azimuthally averaged radial velocity \(V_r\) at some radii inside corotation.

\subsection{Concluding remarks}

The above comparisons lead us to two main conclusions of the discussion on bar pattern speed. (1) The observed properties of relatively isolated SAB and SB galaxies are generally consistent with the simulations of non-interacting galaxies, thereby providing a justification for the \(v_r\)-reversal method described in section \ref{D11} that was motivated by the density wave theory and early numerical works. From its simplicity and good agreement with previous studies, we suggest that the method can be applied to modest-resolution (1 kpc) wide-field velocity field data that are often more easily acquired in large galaxy surveys (such as COMING) compared to high-resolution images. In order to apply the \(v_r\)-reversal method, it is essential to detect CO or another tracer of cold gas (e.g., HI) to radii outside the bar at resolution high enough to resolve the basic galactic structure. (2) The effects of rotation curve (mass distribution) differences and tidal interactions on the location of corotation radius, as shown by simulations, should be taken into account when methods based on velocity (phase) reversals are applied in such systems. Our work encourages new studies, that can be conducted using facilities such as Atacama Large Millimeter/submillimter Array, to further investigate molecular gas kinematics (e.g., radial velocity \(v_r\)) in a variety of morphological and tidal interaction stages and its relation with the bar pattern speed. Establishing a reliable method of determining \(\Omega_\mathrm{b}\) at 1 kpc resolution for a variety of galactic morphologies is especially important when such measurements are attempted for high-redshift objects to understand the evolution of barred galaxies in the early Universe.

\section{Summary}\label{E}

We have presented a mode analysis of the 1-kpc resolution \(^{12}\)CO (1-0) velocity fields of 20 nearby galaxies selected from the COMING legacy project of Nobeyama Radio Observatory. The main results are summarized below.

\begin{enumerate}

\item The velocity fields (moment 1 images) of 13 non-barred (SA) and 7 barred (SAB and SB) spiral galaxies were Fourier decomposed into circular (\(v_\varphi\)) and radial (\(v_r\)) velocity components by ellipse fitting using the program Kinemetry. The fitting results yielded new measurements of the kinematic position angle, disk inclination, and systemic velocity.

\item In barred galaxies, the ratio of noncircular to circular velocity components is of the order of \(\sim0.25\) at inner radii in the bar region, in contrast to an approximately radius-independent value of \(\sim0.15\) in non-barred (SA) spirals. On average, the ratio attains a maximum at \(R/a_\mathrm{b}\sim0.3\). The enhanced noncircular motions are found to be caused by substantial radial velocities, i.e., streaming motion in the bar region with amplitudes up to \(|v_r|\sim40~\mathrm{km~s^{-1}}\) at 1 kpc resolution.

\item The radial velocity in SAB and SB galaxies generally decreases to \(v_r\approx0\) at a reversal radius \(R_\mathrm{r}\), which is found to be \(\mathcal{R}=0.8\mathrm{-}1.6\) the size of the bar in a sample of 7 galaxies. Associating \(R_\mathrm{r}\) with the corotation radius of the bar, we estimated the bar pattern speed of the sample galaxies. The results are consistent with previous studies and suggest that intermediate (SBb-SBc), luminous barred spiral galaxies host fast and slow rotator bars.

\item The bar pattern speed is found to decrease with \(a_\mathrm{b}\) and total galactic stellar mass \(M_\ast\). Although more data are needed to improve the significance of the analysis, the result is in agreement with theoretical predictions that bars evolve by growing and slowing down their rotation as a consequence of angular momentum transfer.

\item Recent numerical simulations of barred galaxies, that include the effects of different galactic mass distributions and tidal interactions, suggest that corotation radius \(R_\mathrm{CR}\) is often located in the vicinity of the \(v_r\)-reversal radius \(R_\mathrm{r}\) (where \(v_r\approx0\)) in relatively isolated galaxies, as assumed in the method to derive \(\Omega_\mathrm{b}\). However, we also find that \(R_\mathrm{CR}\) can be largely offset from \(R_\mathrm{r}\) in galaxies undergoing tidal interactions, suggesting that the velocity reversal method of determining the bar pattern speed in such systems may suffer from large uncertainties.
\end{enumerate}

\begin{ack}
The authors thank the referee for sending valuable comments and suggestions. We are grateful to the staff of Nobeyama Radio Observatory for providing generous help in our observations with the 45-m telescope. The Nobeyama 45-m radio telescope is operated by Nobeyama Radio Observatory, a branch of National Astronomical Observatory of Japan. This research has made use of the NASA/IPAC Extragalactic Database, which is operated by the Jet Propulsion Laboratory, California Institute of Technology, under contract with the National Aeronautics and Space Administration.
\end{ack}

\appendix
\section*{Circular and noncircular velocity components of individual galaxies}

The velocities \(v_\varphi\), \(\Omega\), \(v_r\), and \(\Delta\), derived by mode analysis for all galaxies in the sample (table \ref{tab:gal}) are plotted in figures \ref{fig:plots1}-\ref{fig:plots5}, except for NGC 2903 and NGC 5055, which are shown in figure \ref{fig:ex}. For the definitions of the parameters, see section \ref{C}.

\begin{figure}
\begin{center}
\includegraphics[width=13cm]{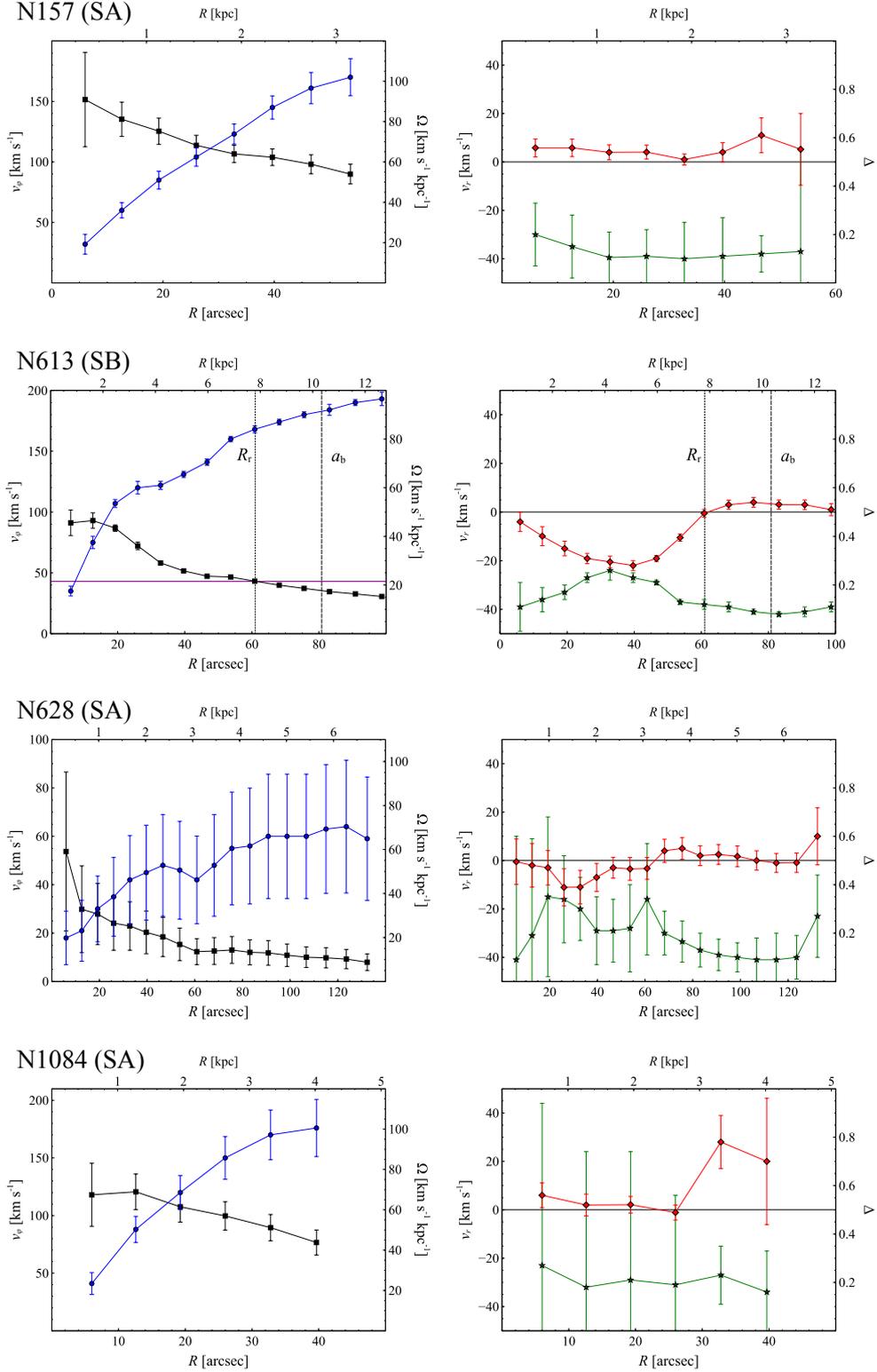}
\end{center}
\caption{Left panels: circular velocity \(v_\varphi\) (blue circles) and angular velocity \(\Omega\) (black squares). Right panels: radial velocity \(v_r\) (red diamonds) and parameter \(\Delta\) (green stars). Bar radius \(a_\mathrm{b}\) and \(v_r\)-reversal radius are indicated by vertical dashed and dotted lines, respectively. The radial velocity \(v_r=s_1/\sin{|i|}\) is not corrected for the near/far side orientation of the galactic disk.}
\label{fig:plots1}
\end{figure}

\begin{figure}
\begin{center}
\includegraphics[width=13cm]{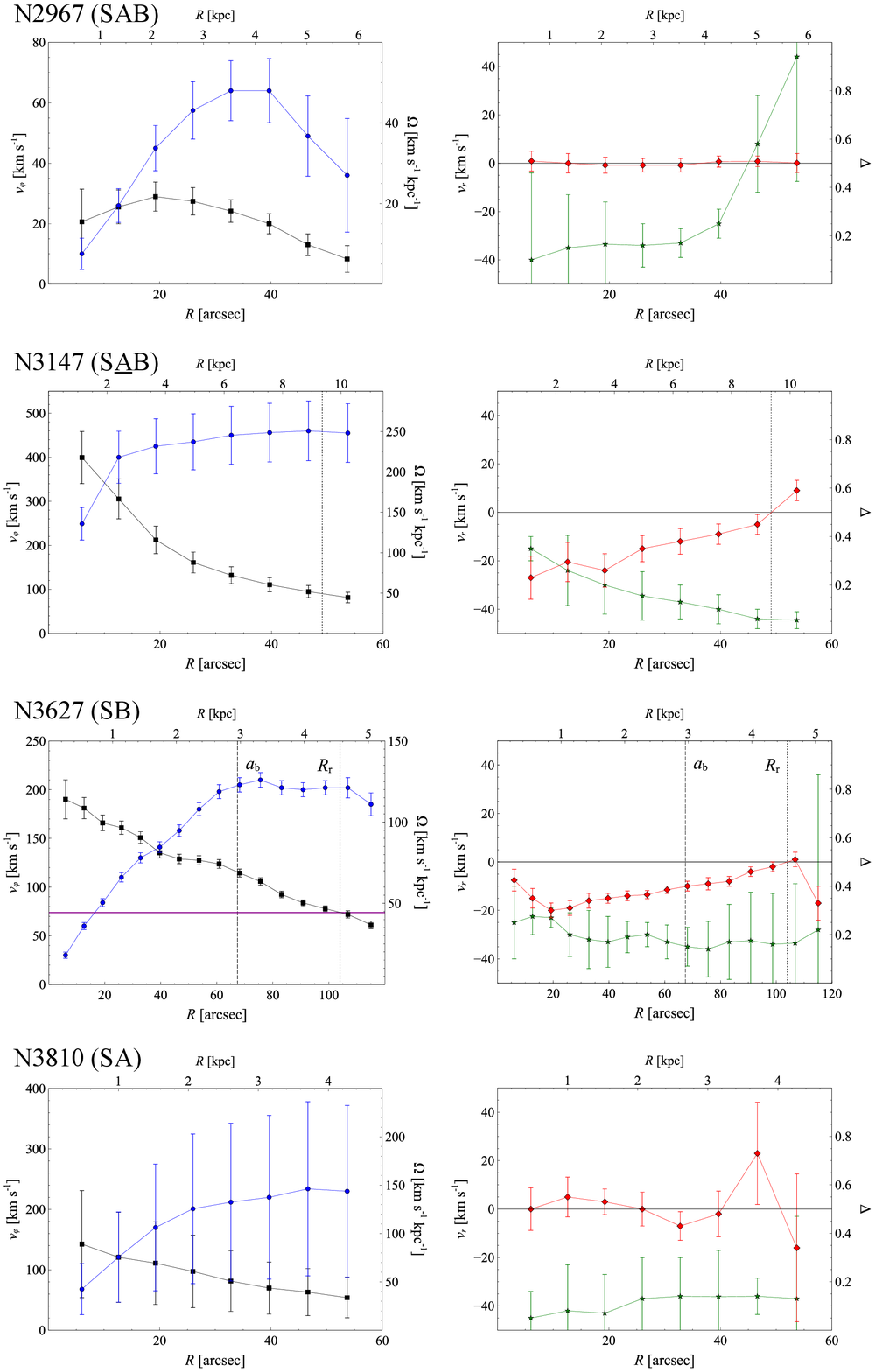}
\end{center}
\caption{Continued (same as figure \ref{fig:plots1}).}
\label{fig:plots2}
\end{figure}

\begin{figure}
\begin{center}
\includegraphics[width=13cm]{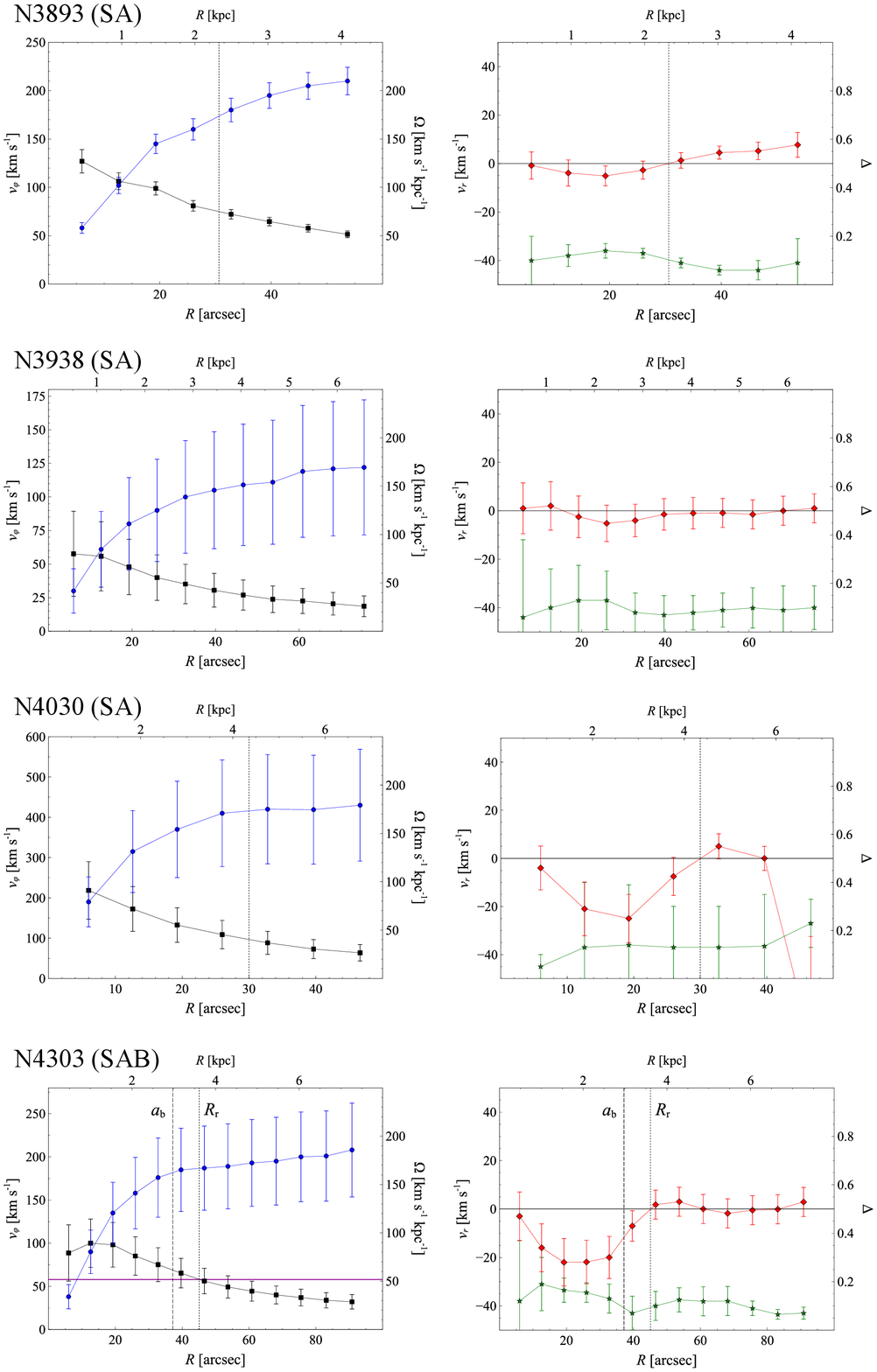}
\end{center}
\caption{Continued (same as figure \ref{fig:plots1}).}
\label{fig:plots3}
\end{figure}

\begin{figure}
\begin{center}
\includegraphics[width=13cm]{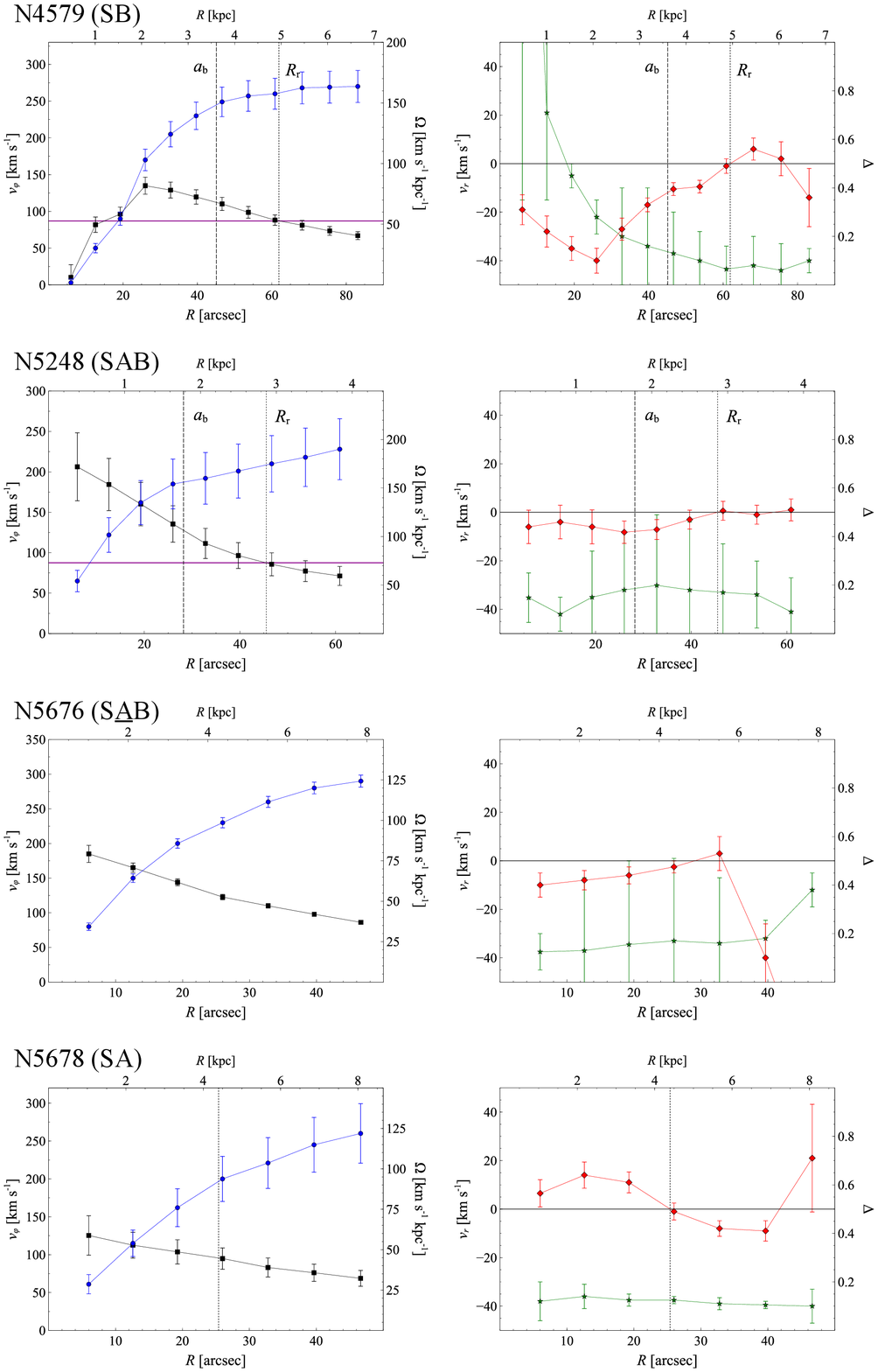}
\end{center}
\caption{Continued (same as figure \ref{fig:plots1}).}
\label{fig:plots4}
\end{figure}

\begin{figure}
\begin{center}
\includegraphics[width=13cm]{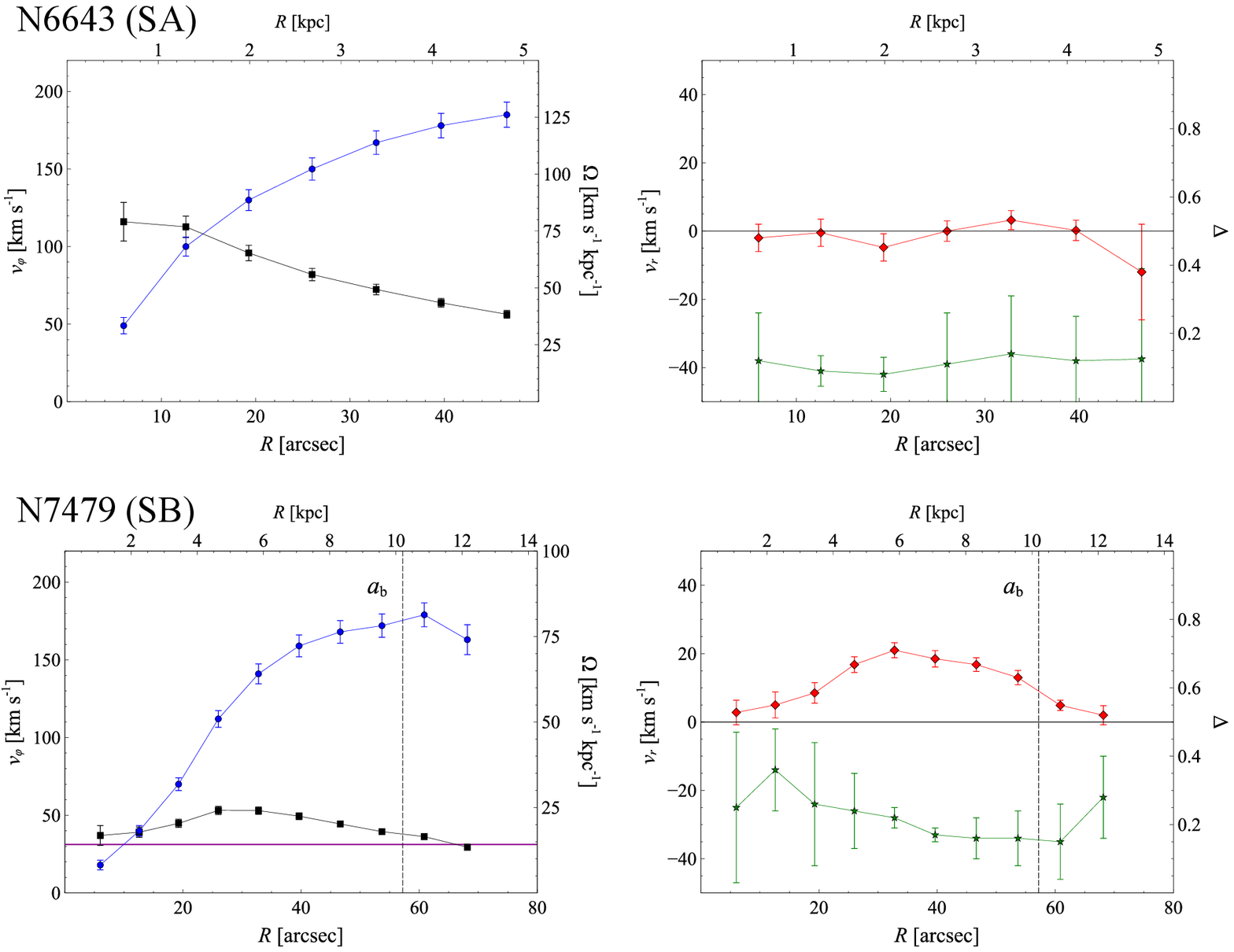}
\end{center}
\caption{Continued (same as figure \ref{fig:plots1}).}
\label{fig:plots5}
\end{figure}

\end{document}